\documentclass[12pt]{article}
\pdfoutput=1
\usepackage[T1]{fontenc}
\usepackage{amsmath,amssymb,mathrsfs}
\usepackage{paralist}
\usepackage{slashed}
\usepackage{graphicx}
\usepackage{subfigure}
\usepackage{color}
\usepackage{fullpage}
\usepackage{hyperref}
\usepackage{amscd}
\usepackage{xcolor}
\usepackage{braket}
\usepackage{amsfonts}
\usepackage{tikz}
\usepackage{cite}
\usepackage{booktabs}
\usepackage{pdfpages}
\usepackage{empheq}
\usepackage{xparse}

\catcode`\@=11 
\def\numberbysection{\@addtoreset{equation}{section} 
        \def\theequation{\thesection.\arabic{equation}}}

\def\be{\begin{equation}} 
\def\ee{\end{equation}} 
\def\ba{\begin{eqnarray}} 
\def\ea{\end{eqnarray}} 
\def\bali{\begin{align}}
\def\eali{\end{align}}

\def\Z{\mathbb{Z}} 
 
\def\RR{\mathbb{R}} 

\def\nl{\nonumber \\}

\def\de{\partial} 
\def\wt{\widetilde}

\def\dag{\dagger}

\def\G{\Gamma} 
\def\D{\Delta} 
\def\d{\delta} 
 
\def\eps{\varepsilon}

\def\l{\lambda}

\def\n{\nu}

\def\r{\rho}

\def\vf{\varphi}

\def\W{\Omega} 
 
\def\th{\theta}

\def\winf{W_{\infty}} 
\def\u1{\widehat{U(1)}}

\begin{document} 
 
\begin{titlepage} 
\begin{center} 
\vskip .6 in 
{\LARGE W-infinity Symmetry in the Quantum Hall Effect }\\
\medskip
{\LARGE Beyond the Edge} 
\vskip 0.2in 
Andrea CAPPELLI${}^{(a)}$ and Lorenzo MAFFI${}^{(b,a)}$
 \medskip

{\em ${}^{(a)}$INFN, Sezione di Firenze\\
Via G. Sansone 1, 50019 Sesto Fiorentino - Firenze, Italy}\\
{\em ${}^{(b)}$Dipartimento di Fisica, Universit\`a di Firenze\\ 
Via G. Sansone 1, 50019 Sesto Fiorentino - Firenze, Italy} \\
\end{center} 
\vskip .2 in 

\begin{abstract}
  The description of chiral quantum incompressible fluids by the
  $\winf$ symmetry can be extended from the edge, where it encompasses
  the conformal field theory approach, to the non-conformal bulk.  The two
  regimes are characterized by excitations with different sizes, energies and
  momenta within the disk geometry.  In particular, the bulk
  quantities have a finite limit for large droplets.  We obtain
  analytic results for the radial shape of excitations, the edge
  reconstruction phenomenon and the energy spectrum of density
  fluctuations in Laughlin states.
\end{abstract} 
 
\vfill 
\end{titlepage} 
\pagenumbering{arabic} 
\numberbysection

 
\section{Introduction}

A long-standing and long-term problem in the quantum Hall effect
\cite{gp} is the understanding of Laughlin incompressible fluids
\cite{laughlin}, their geometry and dynamics, in relatively simple
analytic terms.  For what concerns the edge physics, a clear
description has been given by the conformal field theory (Luttinger
liquid) approach \cite{wen-book}. Conformal theories has been
developed and generalized to a great extent, leading to extensive
model buildings \cite{pfaff}\cite{cv}. Their bulk counterparts are
expressed by the topological Chern-Simons gauge theory, complemented
by the Wen-Zee terms accounting for geometric responses
\cite{wen-zee}.  All these results are analytic and exact in the
low-energy limit and give access to universal properties of Laughlin
and Jain states.

The study of low-energy bulk physics above the topological limit is
somehow less developed. The Jain theory of flux attachment and
composite fermion excitations has been confirmed experimentally and
numerically \cite{jain}, and it remarkably extends to relativistic
regimes for filling fractions near one half \cite{son-rel}.  Flux
attachment has been successfully described by mean field theory
\cite{hlr}\cite{fradkin} and Hartree-Fock approximations
\cite{shankar}.  However, the nature of the composite fermion
excitation remains rather unclear.

Other approaches have been trying to develop physical models
of the composite fermion: one idea is that this is a dipole made by a
fermion (elementary) bound to a hole/vortex (collective), whose
charges do not precisely cancel each other \cite{hp} \cite{read}
\cite{camilla}.
In hydrodynamic
approaches, the composite fermion is associated to the slow motion of
vortex centers, rather than the fast motion of the fluid itself \cite{hydro}.
These studies have also suggested that additional degrees of
freedom may be needed, such as a two-dimensional dynamic
metric \cite{haldane} \cite{bimetric}.

Another interesting result has been the derivation of the magneto-roton minimum
(several minima in general) by a rather simple analysis of fluctuations
of the Fermi surface near half filling \cite{son-mr}.
This work suggested that the minima occur at momentum values that are
universal, i.e. invariant under deformations of the Hamiltonian
within the gapped phase.

These theoretical advances suggest a number of questions that
motivate the present work:
\begin{itemize}
\item
  Does the dynamics of quantum incompressible fluids encompass all the
  bulk physics of (spin polarized) Laughlin and Jain states, or
  additional features/degrees of freedom should be introduced?
\item
  Is it sufficient to restrict oneself to the lowest Landau level
  for Laughlin filling fractions $\nu=1/(2s+1)$?
\item
  Given that excitations of incompressible fluids are described by the
  $\winf$ symmetry \cite{ctz1}\cite{sakita}\cite{read},
  is the relevant physics all accounted for by the
  representations of this algebra?  
\item
  Are there universal features in the low-energy sector(s) of
  bulk excitations?
\end{itemize}

In this paper, we show that the analytic methods of
the $\winf$ symmetry, i.e. the study of its algebra and
representations \cite{ctz-class},
can be extended from the domain of edge physics to the description
of bulk dynamics.
Our methods are introduced in the filled lowest Landau level and then
extended to fractional fillings by bosonization.
We consider the geometry of the disk with radius $ r< R$, that
is suitable for discussing both edge and bulk excitations.
We first show that the $\winf$ algebra is nicely rewritten in terms
of the Laplace transform of the two-dimensional electron density
with respect to $r^2$. This quantity is actually the generating function
of the higher-spin generators in the algebra, whose expectation values
express the radial moments of the density.

The $\winf$ algebra of Laplace transformed densities takes a simple form
in the limit to the
edge. As formulated in Ref.\cite{cm}, this amounts to the combined expansion
of radii $r$ and angular momenta $m$ for values near the edge of the
droplet, as follows:
\be
r=R+x,\qquad m=\frac{R^2}{\ell^2}+m',\qquad |x|<\ell,
\qquad |m'|<\frac{R}{\ell}, \qquad R\to\infty,
\label{edge-lim}
\ee
where $\ell$ is the magnetic length. In this limit, we reobtain the conformal
field theory setting, and show that $\winf$ representation theory allows to
compute the radial profile of edge excitations, for integer and fractional
fillings.

Next, we observe that our approach can be
extended to `large' fluctuations, identified by
parameter values,
\be
x=O(\ell),\qquad m'=O\left(\frac{R}{\ell}\right),
\qquad {\bf k}\;\ell=\frac{m'\ell}{R}=O(1),
\qquad {\bf p}\;\ell =O(1) .
\label{edge-large-lim}
\ee
corresponding to finite $R\to\infty$ limits of edge ({\bf k}) and bulk
({\bf p}) momenta, respectively parallel and orthogonal to the edge.
More importantly, the form of the Hamiltonian for short-range two-body
potentials can be found in this regime, where it admits a
bosonic form, allowing analytic computations of the excitation
spectrum.

Within this nonlinear, non-conformal setting, we can analyze both bulk
and boundary fluctuations and obtain the following results:
\begin{itemize}
\item
For large excitations, the spectrum is nonlinear and shows a
minimum at finite momenta ${\bf k}\;\ell=O(1)$.
This is the so-called edge reconstruction phenomenon  \cite{edge-reco} (also
called `edge roton' minimum \cite{jain-rot}),
namely the tendency of the electron droplet to
expel a thin shell at finite radial distance, for shallow confining
potentials.
\item
  Particle-hole excitations of the flat density possess a
  monotonic growing energy with respect to bulk momentum $({\bf p})$; instead,
  fluctuations around deformed densities, made by accumulating a large charge
  at the edge, show an oscillating spectrum.
\end{itemize}
These results illustrate a new analytic method for
understanding the low-energy bulk dynamics of quantum incompressible fluids.
The study of excitations presented in this paper is not exhaustive
and thus cannot exactly pinpoint the magneto-roton excitation
seen in experiments and simulations \cite{mag-rot}.
Nevertheless, more in-depth analyses of this approach may
identify this feature.

 Our results suggest that the answer to the four questions formulated
 earlier could be affirmative. The study of $\winf$ representations in
 the extended range surely provides universal low-energy bulk
 features, because the deformation of conformal theory is uniquely
 determined.  The flux attachment is automatically accounted for by
 generalizing the bosonic expressions from $\nu=1$ to
 Laughlin states $\nu=1/(2s+1)$, $s=1,2,\dots$.
 Finally, the restriction to the lowest
 Landau level seems to be correct in this case.

The outline of the paper is the following.  In Section two, we briefly
recall previous results of the $\winf$ symmetry approach.  The
$\winf$ algebra has been known since the early days of quantum
Hall physics \cite{gmp}. It has been fully understood within the conformal
field theory of edge excitations \cite{ctz-class}, where it identifies special
`minimal models' that precisely match the Jain states \cite{ctz-min}.
Moreover, it implements corrections to bosonization 
beyond Luttinger theory \cite{ctz-dyna}.
In Section three, we study the algebra of Laplace transformed densities
in the edge limit (\ref{edge-lim}) and obtain the radial profile
of excitations.
In Section four, we analyze the algebra and two-body Hamiltonian in the
second limit of large excitations (\ref{edge-large-lim}) and obtain
the results for bulk physics: the edge reconstruction and edge roton
effects and the energy spectrum of bulk fluctuations in presence
of a large boundary charge.

\section{$W_\infty$ symmetry of quantum Hall
  incompressible fluids}

\subsection{Quantum area-preserving diffeomorphisms}

\begin{figure}[h]
\begin{center}
\includegraphics[width=0.7\textwidth]{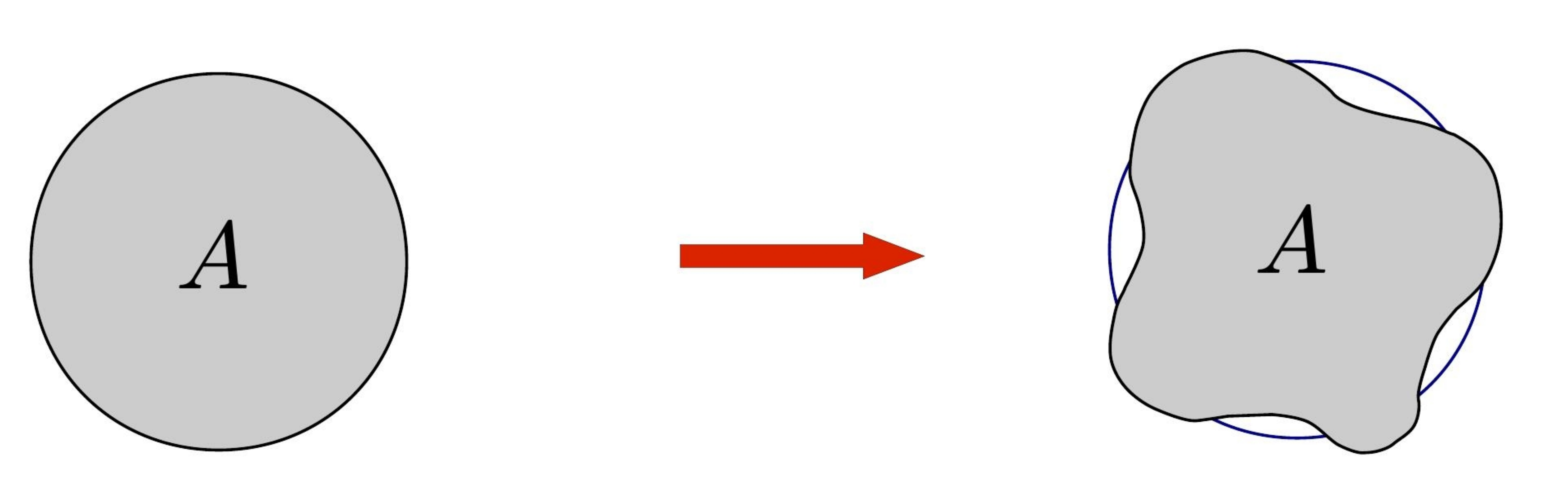}
\caption{Shape deformation of the droplet 
under the action of area preserving diffeomorphisms.}
\label{fig-adiff}
\end{center}
\end{figure}

A two-dimensional incompressible fluid is characterized at the
classical level by a constant density $\r_o$.
Suppose that the fluid is confined inside a rotational invariant
potential, such that it forms a round droplet 
with sharp boundary (see Fig. \ref{fig-adiff}).
Density fluctuations are forbidden in the bulk, thus excitations
amount to shape deformations of the droplet. They keep
the same area ${\cal A}$, because the
number $N$ of constituents of the fluid, say electrons, is also constant,
$N=\r_o\mathcal{A}$.

We can generate the deformed droplets by coordinate
transformations of the plane that keep the area constant:
these are the so-called two-dimensional
area-preserving diffeomorphisms \cite{ctz1}\cite{sakita}.
Their action can be expressed in terms of Poisson brackets, in analogy
with canonical transformations of a two-dimensional phase space.
It is convenient to use complex coordinates of the plane,
\be
z=x^1+ix^2, \qquad \bar z=x^1-ix^2,\qquad ds^2=dz d\bar z,\qquad
\d_{z\bar z}=\frac{1}{2}, \qquad \d^{z\bar z}=2.
\label{z-zbar}
  \ee
The Poisson brackets are:
\be
\{f,g\}=\eps^{z\bar z}\de_z f\;\de_{\bar z} g +
\eps^{\bar z z}\de_{\bar z} f\;\de_z g, \qquad
\eps^{z\bar z}= -\eps^{\bar z z}=-2i .
\label{pb}\ee
The area-preserving transformations are given by,
\be
\d_w z=\{z,w(z,\bar z)\},
\label{w-transf}
\ee
in terms of the generating function $w(z,\bar z)$.
The deformations of the ground-state density $\r_{GS}$
follows from the chain rule:
\be
\r_{GS}=\r_o \Theta\left(R^2-z\bar z\right),\qquad
\d_w\r_{GS}=\{\r,w\}\propto \d \left(z\bar z-R^2\right),
\label{c-fluct}
\ee
where $\Theta$ is the step function. We see that
fluctuations have support at the boundary $\bar z z =R^2$, as expected.

A basis of generators can be obtained by expanding the function
$w(z,\bar{z})$ in power series,
\be
\mathcal{L}_{n,m}=z^{n+1}\bar{z}^{m+1},\qquad
w(z,\bar{z})=\sum_{n,m\geq -1}c_{nm}z^{n+1}\bar{z}^{m+1}.
\label{w-gen}
\ee
These generators obey the following algebra:
\be
\label{winfty}
\left\lbrace\mathcal{L}_{n,m},\mathcal{L}_{k,l}\right\rbrace=
\left[\left(m+1\right)\left(k+1\right)-
\left(n+1\right)\left(l+1\right)\right]\mathcal{L}_{n+k,m+l},
\ee
that is called the $w_{\infty}$ algebra of classical
area-preserving diffeomorphisms.

We now consider the realization of this symmetry in the quantum Hall
effect, starting from the lowest Landau level for simplicity \cite{ctz1}.
In the quantum theory, the generators (\ref{w-gen}) become
one-body operators,
\be
 \hat{\mathcal{L}}_{n,m}  =\int d^2z\ \hat\Psi^\dag(z,\bar z)\ 
 \bar{z}^{m+1} z^{n+1}\ \hat \Psi(z,\bar z),
 \qquad n,m\ge -1,
\label{lnm-def}\ee
where the field operator $\hat{\Psi}(z,\bar{z})$ is,
\be
\hat \Psi(z,\bar z)=e^{-z\bar z/2\ell^2}\ \hat \vf(z),
\qquad \hat \vf(z)= \frac{1}{\ell\sqrt{\pi}}
\sum_{n=0}^\infty\left(\frac{z}{\ell}\right)^n\frac{1}{\sqrt{n!}}\ 
\hat c_n,\qquad \{\hat c_n,\hat c^\dag_m\}=\d_{n,m}.
\label{field-op}
\ee
In this expression, there appear the analytic field $\hat\vf(z)$, the
fermionic creation-annihilation operators,
$\hat c_n,\hat c^\dag_m$, and the magnetic length $\ell=\sqrt{2 \hbar/B}$.
We set $\ell=1$ in the following, unless when specified.

The coordinates become non-commuting, $[\bar z,z]=\ell^2$, when acting
on analytic fields $\hat \vf (z)$, and classical polynomials in
$z,\bar z$ require a choice of normal ordering.
The definition of $\hat{\mathcal{L}}_{n,m}$ in (\ref{lnm-def}) adopts
the standard prescription, setting $\bar z $ to the left of
$z$ and then replacing $\bar z \to \ell^2\de_z$, the holomorphic
derivative.  Other normal-ordering choices are possible:
in the following, it is convenient to consider the case of anti-normal
ordering, placing $\bar z$ to the right of $z$. In this case, the
generators read:
\ba
\hat{\mathcal{L}}_{n,m} &=&
\int d^2z\ e^{-z\bar z/\ell^2}\ \hat{\vf}^\dag(z)\ 
z^{n+1}\left(\ell^2\de_z\right)^{m+1} \hat \vf(z)
\nl
&=& \ell^{n+m+2}\sum_{k\ge 1}\frac{\sqrt{(k+m)!(k+n)!}}{(k-1)!}\
c^\dag_{k+n}c_{k+m} .
\label{lnm-expr}
\ea
The Fock-space expression is useful to derive 
the following commutation relations,
\be
\begin{aligned}
  \left[\hat{\mathcal{L}}_{n,m},\hat{\mathcal{L}}_{k,l}\right]=
  \sum_{s=1}^{Min(m,k)}
  &\frac{\ell^{2s}\ \left(m+1\right)
    !\left(k+1\right)!}{\left(m-s+1\right)!\left(k-s+1\right)!s!}
  \hat{\mathcal{L}}_{n+k-s+1,m+l-s+1}\\
&-\left(m\leftrightarrow l,n\leftrightarrow k\right) ,
\end{aligned}
\label{Winfty}
\ee
that are called the $W_{\infty}$ algebra of quantum area-preserving
transformations.

Let us discuss a number of properties of this algebra \cite{ctz1}.
\begin{itemize}
\item
  On the right hand side there occur a finite number of terms involving
  increasing powers of $\ell^2=2\hslash/B$; the first term corresponds to the
quantization of the classical algebra \eqref{winfty}, the others
are higher-order quantum corrections; their form changes for
other normal-orderings of $\hat{\mathcal{L}}_{n,m}$.
\item
  The $\winf$ algebra contains an infinite number of Casimir invariants,
  the operators $\hat{\mathcal{L}}_{n,n}$ that actually correspond to
  radial moments $O(r^{2n+2})$ of the density (cf. (\ref{lnm-def})).
  In particular $\hat{\mathcal{L}}_{0,0}=z \de_z$ is the angular momentum
  operator and obeys the algebra:
  \be
  \left[\hat{\mathcal{L}}_{0,0},\hat{\mathcal{L}}_{n,m}\right]=
  (n-m) \hat{\mathcal{L}}_{n,m} .
  \label{l-zero}
  \ee
\item
  It is apparent from the second-quantized expression (\ref{lnm-expr}) that the
  $\hat{\mathcal{L}}_{n,m}$ create bosonic particle-hole excitations
  with angular momentum jump $\Delta k=n-m$, whose amplitude is parameterized
  by another integer, say $m$.
\item
  Area-preserving diffeomorphisms can be similarly implemented in
  the higher Landau levels, by replacing the corresponding field
  operators in (\ref{lnm-def}). Actually, the transformations act within each
  level.  The second quantized expression (\ref{lnm-expr}) is valid for
  any level by just replacing the corresponding creation-annihilation
  operators \cite{ctz1}.
\item
  The $\winf$ algebra can be written in another basis by expressing
  the generating function (\ref{w-gen}) in terms of plane waves, 
  $ w_{k,\bar k}(z,\bar z) =\exp(ik\bar z/2+i\bar k z/2)$.
  Owing to Eq.(\ref{lnm-def}), the corresponding quantum generators become
 the Fourier modes of the density $\hat \r(k,\bar k)$,
  obeying the Girvin-MacDonald-Platzman algebra \cite{gmp}:
  \be
  \left[\hat \r(k,\bar k), \hat \r(p,\bar p)\right]=
\left(e^{p \bar k/4}- e^{\bar p k/4} \right)
  \hat \r(k+p,\bar k+ \bar p) .
  \label{gmp-alg}\ee
  It is apparent that the $\hat \r(k,\bar k)$ corresponds to
  the generating function of polynomial operators $\hat{\mathcal{L}}_{n,m}$
  (in the so-called Weyl ordering, e.g.
  $z \bar z\to (z\de/\de z +\de/\de z\; z)/2$). The Fourier basis is
  better known in the literature, but the polynomial basis is more
  convenient for the following discussion of excitations.
\end{itemize}

Let us now find the action of $\winf$ generators on the ground state.
Consider the lowers Landau level completely filled by $N$ electrons.
Its second quantized expression is:
\be
\vert \W\rangle =\vert N, \nu=1\rangle =
c^\dag_{N-1}c^\dag_{N-2}\cdots c^\dag_1c^\dag_0\vert 0\rangle .
\label{gs}\ee
Being a completely filled Fermi sea, this state does not admit
particle-hole transitions decreasing the total angular momentum,
that are generated by $\hat{\mathcal{L}}_{n,m}$ with $\Delta k =n-m<0$,
see Eq.(\ref{lnm-expr}). It then follows that the ground state obeys
the following conditions:
\be
\hat{\mathcal{L}}_{n,m}\vert\W\rangle =0, \qquad
    {\rm for}\qquad -1\le n< m .
    \label{hws-int}
\ee
These are called highest-weight conditions, borrowing the language of 
infinite-dimensional algebra representations \cite{cft}.
The highest-weight state
is the top (or bottom) state of an infinite tower, and
is annihilated by half of the ladder operators.

The action of the other $\winf$ generators is the following.
The Casimirs leave the ground state invariant and possess eigenvalues
that are polynomials in $N$. For example, the operators,
\be
\hat{\mathcal{L}}_{-1,-1}\vert\W\rangle = N \vert\W\rangle ,\qquad
\hat{\mathcal{L}}_{0,0}\vert\W\rangle = \frac{N(N-1)}{2} \vert\W\rangle,
\label{q-m}\ee
respectively measure the number of particles and the angular momentum.

The $\winf$ generators with $\Delta k =n-m>0$ increase the angular momentum
and create excitations:
\be
\hat{\mathcal{L}}_{n,m}\vert\W\rangle =\vert {\rm excit}\rangle, \qquad
    {\rm for}\qquad  n> m\ge -1 .
    \label{w-excit}
\ee
Actually, it can be shown that they generate the entire space of
neutral excitations in the lowest Landau level, because the
relation (\ref{lnm-expr}) between $\hat{\mathcal{L}}_{n,m}$ and
fermionic bilinears is invertible.  These generators provide a
kind of non-relativistic bosonization of Hall electrons \cite{ctz1}.

In conclusion, we have shown that the $\winf$ transformations
are spectrum-generating, i.e. correspond to the dynamical symmetry
of the $\nu=1$ quantum Hall system. Let us add some remarks.
\begin{itemize}
\item
  As is well known, the algebraic approach of dynamical symmetries
  is only useful when the Hamiltonian can be simply written in terms of
  the generators. Later we shall see that short-range two-body
  interactions can indeed be included.  So far, we only considered
  free electrons in a confining potential: a quadratic form
  $V(z,\bar z)=\l \bar z z$ actually corresponds to the generator
  $\hat{\mathcal{L}}_{0,0}$ with simple commutation relations (\ref{l-zero}).
\item
 Different normal orderings of the $\winf$ generators amount to
  redefinitions,
  \be
  \hat{\mathcal{L}}_{n,m}\quad \longrightarrow \quad
  \hat{\mathcal{L}}_{n,m}+ a \hat{\mathcal{L}}_{n-1,m-1}+
  b \hat{\mathcal{L}}_{n-2,m-2} + \cdots ,
\label{normal-ord}  \ee
where $a,b,\cdots$ are numerical constants, as easily seen by
Eq.(\ref{lnm-def}).
  It follows that the ground-state
  conditions (\ref{hws-int}) are valid for any choice of normal ordering.
\end{itemize}

\subsection{W-infinity symmetry of Laughlin states}

The Laughlin ground states with fillings $\nu=1/(2s+1)=1/3,1/5,\dots$
are better discussed in coordinate representation.
Let us rewrite the $\winf$ generators introduced earlier for $\nu=1$
in terms of the electron coordinates $z_i$, $i=1,\dots,N$. They read,
\be
\hat{\mathcal{L}}_{n,m}=\sum_{i=1}^N z_i^{n+1} \; \de_{z_i}^{m+1} .
\label{lnm-coord}\ee
The (analytic part of the) ground state wave function is written
\be
\Psi_1(z_1,\dots, z_N) =\D\left(\left\{z_i\right\}\right) ,
\qquad\quad \D\left(\left\{z_i\right\}\right)=\prod_{1=i<j=N}
\left(z_i-z_j \right) ,
\label{gs-coord}\ee
in terms of the Vandermonde determinant.

The $\winf$ highest-weight conditions (\ref{hws-int}) are rewritten:
\be
\hat{\mathcal{L}}_{n,m}\Psi_1=
\sum_{i=1}^N z_i^{n+1} \; \de_{z_i}^{m+1}
\prod_{1=i<j=N} \left(z_i-z_j \right)=0, \qquad -1\le n<m .
\label{hws-coord}\ee
Note that these conditions are not completely trivial in
terms of coordinates and actually involve some polynomial identities.

We now define new $\winf$ generators that fulfill the same
highest-weight conditions when acting on Laughlin states.
The Laughlin wavefunction reads, 
\be
\Psi_{2s+1}(z_1,\dots, z_N) =\D\left(\left\{z_i\right\}\right)^{2s+1},
\qquad\quad \nu=\frac{1}{2s+1} .
\label{laugh-wf}\ee
We modify the $\winf$ generators 
by performing the similarity transformation \cite{flohr}, 
\ba
\hat{\mathcal{L}}_{n,m}^{(1)} \qquad  \longrightarrow \qquad
\hat{\mathcal{L}}_{n,m}^{(2s+1)}
&=& \D\left(\left\{z_i\right\}\right)^{2s}\ \hat{\mathcal{L}}_{n,m}^{(1)} 
\ \D\left(\left\{z_i\right\}\right)^{-2s}
\nl
&=& \sum_{i=1}^N z_i^{n+1}
\left(\de_{z_i} -\sum_{j,j\neq i}\frac{2s}{z_i-z_j}\right)^{m+1} .
\label{lnm-laugh}\ea
The new generators $\hat{\mathcal{L}}_{n,m}^{(2s+1)}$ clearly obey the
highest-weight conditions (\ref{hws-int}) on Laughlin wavefunctions: for $n<m$,
they cannot compress it (although some other operators could);
for $n>m$, they generate particle-hole excitations above this state.

Other nice features are:
\begin{itemize}
\item
  The operators (\ref{lnm-laugh}) fulfill the same $\winf$ algebra
  (\ref{Winfty}), owing to the similarity transformation.  As discussed in
  Ref.\cite{flohr}, this rather singular transformation may lead to
  non-holomorphic delta-function singularities, but these are irrelevant in
  expectation values, due to the zeroes of wavefunctions.
\item
  It can be shown that the modified derivatives in (\ref{lnm-laugh}) are
  actually arising by coupling electrons to a Chern-Simons statistical field
  with coupling constant $1/2s$, that ``attaches and even number of flux
  quanta to electrons'' \cite{fradkin}. Therefore the similarity transformation
  (\ref{lnm-laugh}) is actually the Jain composite-fermion map between integer
  and Laughlin Hall states \cite{jain}.
\end{itemize}

\subsection{$\winf$ symmetry of edge excitations}

The results of the previous sections can be summarized by saying that
integer and fractional Hall states possess the dynamical symmetry of
quantum area-preserving diffeomorphisms.
Given that these generators span the space of excitations,
they should account for both bulk and edge properties
of the quantum incompressible fluid.

However, few results were obtained in this general setting because
the $\winf$ generators are too singular in the $N\to\infty$
limit of large droplets, that is relevant for universal features.
The eigenvalues of the Casimirs (\ref{q-m}) grow polynomially in $N$,
the Fock space expression (\ref{lnm-expr}) and the covariant
derivatives (\ref{lnm-laugh})
require regularization and renormalization. As a consequence, the
$\winf$ algebra (\ref{Winfty}) does not have a finite
large $N$ limit and its renormalized form was not understood so far.

Fortunately, it was found that a well-defined formulation of the
$\winf$ symmetry exists in the Hilbert space of $(1+1)$-dimensional
massless field theories, that describe the edge excitations of the Hall
droplet \cite{ctz-class}.
In this setting, $\winf$ is an extension of the conformal
symmetry  and the corresponding representations were
completely understood in the mathematical literature \cite{kac}.
The $\winf$ algebra on the circle edge has a well defined
large $N$ limit, where it acquires a unique central extension,
corresponding to the conformal anomaly \cite{cft}.

Therefore, the works \cite{ctz-class}\cite{ctz-min}
proposed to study the $\winf$ symmetry directly in the
edge theory and to understand its implications for the conformal
field theory description.

The edge excitations of the filled lowest Landau level are described
by a massless chiral (Weyl) fermion in one dimension \cite{cdtz}:
both theories possess the same one-component
fermionic Fock space, but amplitudes and measure of integration are
different.  The edge is identified as the circle $C_R$, 
$ z=R\exp(i\th)$,  and the radial dependence is lost. The earlier quantization
(\ref{lnm-def}) of $w_\infty$ generators is replaced by the
following expression:
\be
\hat{\mathcal{L}}_{n+i,i} \quad \longrightarrow \quad -
\hat{V}_{-n}^{(i+2)}=\oint_{C_R} \frac{dz}{i z}\ \hat\psi^\dag(z)\
z^n \left(z\de_z\right)^{i+1} \hat\psi(z),
\label{lnm-edge}\ee
involving the Weyl fermion field,
\be
\hat\psi(z)=\frac{1}{\sqrt{2\pi}}
\sum_{k=-\infty}^{\infty} z^{k-1/2} \ \hat c_{k}, \qquad \left\{\hat
  c_n,\hat c^\dag_m\right\}=\d_{n,m}.
\label{weyl-def}\ee
Note that $z\de_z\to -i \de/\de\th$ on the circle and that the $N$
dependence has disappeared.

Therefore, the $\winf$ algebra on the circle is different from
(\ref{Winfty}), with the exception of the leading $O(\hbar)$ term.
Nevertheless, it
still contains infinite Casimirs $\hat V^{(i)}_0$, $i=0,1,2,\dots$
that correspond to conserved charges in the
conformal theory. The operators $\hat{V}_{n}^{(1)}$ are identified as
the modes $\hat\r_n$ of the edge density $\hat\r(z)$ in the Weyl
theory, while $\hat{V}_{n}^{(2)}$ are the Virasoro operator
$\hat{L}_{n}$ \cite{ctz-class}. Their algebra reads (omitting
the hats on operators hereafter):
\ba
&&\left[\r_n,\r_m\right]=n \d_{n+m,0},
  \label{rho-al}\\
  &&\left[L_n,\r_m\right]=-m\r_{n+m},
    \nl
    &&\left[L_n,L_m\right]=(n-m)L_{n+m} +
    \frac{c}{12}n\left(n^2-1 \right)\d_{n+m,0},
    \label{curr-alg}
\ea
where $c$ is the Virasoro central charge ($c=1$ for the Weyl fermion).
The algebra obeyed by higher-spin operators can be found in
\cite{ctz-dyna}.

The ground state
relations (\ref{hws-int}) obeyed by the $\winf$ generators
are mapped in the highest weight conditions of conformal representations,
such as \cite{ctz-class}:
\be
\r_k|\W\rangle=0, \qquad L_k|\W\rangle=0,\qquad k \ge 0.
\label{hws2}\ee

The physical implications of the $\winf$ symmetry have been investigated for
general edge theories beside the Weyl fermion pertaining to $\nu=1$:
here we summarize the results.
\begin{itemize}
\item
  In the simplest case, $\winf$ is the enveloping
  algebra of the current algebra with central charge $c=1$:
  namely, the higher-spin currents correspond to polynomials of the edge
  density $\r(z)$, generalizing the so-called Sugawara construction:
  $V^{(1)}(z)=\r(z)$, $ V^{(2)}=: \r(z)^2:$, $V^{(3)}=:\r(z)^3:$,
  etc \cite{ctz-class}. One recovers the chiral Luttinger theory
  (compactified chiral boson) of Laughlin edge states. The
  higher-spin currents are automatically implemented without any
  further condition on the conformal theory.
\item
  The description of edge excitations for the Jain
  states with $\nu=n/(2s n \pm 1)$ is based on the $n$-component
  current algebra with central charge $c=n$, whose spectrum is
  parameterized by an integer-valued symmetric matrix, called $K$
  matrix.  The specific form of $K$ describing Jain states is obtained from
  simple phenomenological arguments based on the composite fermion
  map.  General $\winf$ representations also correspond to current
  algebras with $c=n$: however, there is a fine structure that
  allows to identify ``minimal models'' of $\winf$ symmetry,
  possessing a reduced set of excitations \cite{ctz-min}. These models
  realize the $U(1)\times SU(n)_1$ extended symmetry and precisely
  select the $K$ matrix for Jain states.  Therefore, the $\winf$
  symmetry and the requirement of minimality are sufficient to
  a-priori identify the Jain states without any phenomenological
  input.  This purely geometric characterization of the prominent Hall
  states is rather remarkable.
\item
  The non-Abelian Hall states, such as the Pfaffian with
  central charge $c=3/2$, cannot directly be identified among
  $\winf$ symmetric theories. The explanation of this fact
  is the that additional features are present. In the case of the Pfaffian,
  this is the electron pairing mechanism, leading to the incompressible
  fluid of bosonic compounds. In the works \cite{cgt}, this pairing
  was described by a projection in a two-fluid $c=2$ model that possess
  the $\winf$ symmetry. This description of the Pfaffian by
  an Abelian ``parent'' state brings back in the $\winf$ symmetry and
  also implies some useful relations for wavefunction modeling.
\item
  While the lower Casimirs $\r_0$ and $L_0$ respectively express
  the charge and Hamiltonian $ H=v L_0/R$ of edge excitations, the
  higher ones encode non-relativistic corrections
  $\D H^{(i)}= V^{(i)}_0/R^{i-1}$, $i=3,4,\dots,$ to the energy spectrum,
  forming a series in $1/R$.
\end{itemize}

In conclusion, the $\winf$ symmetry of quantum incompressible fluids
has definite implications for edge theories.
On the contrary, bulk excitations were not much investigated
beyond the rather formal results presented in Section 2.1
and 2.2.  In particular, the map (\ref{lnm-edge}) between the
bulk operators ${\mathcal{L}}_{n,m}$ (for a given normal ordering)
and the edge generators $V_n^{(i)}$ was unclear.

\section{Radial shape of edge excitations}

\subsection{Laplace transform and bulk-boundary map}

A first understanding of the bulk-boundary map 
was obtained only recently, by a more careful definition of
the edge limit \cite{cm} (See also \cite{azuma}).
It was shown that the modes of the
edge current $\r_n$ are simply obtained by integrating out
the radial dependence of the bulk density,
\be
\r_m = \int_0^\infty dr r \int_0^{2\pi} d\th\; \r(r,\th) e^{-im\th},
\label{rho-int}\ee
and by taking the limit $R\to\infty$ for values of coordinate $r$ and
angular momentum $m$ at the edge of the droplet, as follows:
\be
r=R+x, \qquad |x|<1, \qquad m=R^2+m',\qquad |m'|<R,\qquad R\to\infty .
\label{edge-lim2}\ee
In this limit, the $\r_n$ become operators in the edge conformal theory
and obey the expected current algebra commutation relations,
$\left[\r_n,\r_m\right]=n \d_{n+m,0}$.

In this Section, we are going to extend this map to
higher-spin $\winf$ generators and the full algebra,
thus clearly establishing the link between bulk and edge descriptions.
 Furthermore, in Section four this setup will allow us to go beyond
the conformal theory description.

The key idea is to consider the Laplace transform of the bulk density
with respect to $r^2$:
\be
\r_k(\l)= \int_0^\infty dr r e^{-\l r^2}
\int_0^{2\pi} d\th\; \hat\r(r,\th) e^{-ik\th} .
\label{lap-def}
\ee
Expanding the field operators (\ref{field-op}) in
the Fock basis, we find:
\be
\r_k(\l)= 
\sum_{j=0}^\infty \frac{1}{(1+\l)^{j+1+k/2}}
\frac{\G\left(j+1+\frac{k}{2} \right)}{\left(\G(j+1)\G(j+k+1)\right)^{1/2}}
c^\dag_j c_{j+k} .
\label{lap-alg}
\ee
It turns out that this quantity possesses a finite $R\to\infty$ limit,
up to a multiplicative factor. Taking the limit according to
(\ref{edge-lim2}), we obtain,
\be
\r_k(\l)= \frac{1}{\left(1+\l\right)^{R^2+\mu+1} }
\sum_{j'=-R^2}^\infty (1+\l)^{\mu-j'-k/2} :c_{j'}^\dag c_{j'+k}:,
\qquad R\to\infty .
\label{rho-lim1}
\ee
Upon removing the prefactor, $(1+\l)^{-R^2-\mu-1}$, we obtain a well-defined
quantity in the Weyl-fermion edge theory (\ref{weyl-def}):
\be
\r_k(\l)\ \longrightarrow\ 
\sum_{j'=-R^2}^\infty (1+\l)^{\mu-j'-k/2} :c_{j'}^\dag c_{j'+k}:.
\label{rho-lim}
\ee
In these expression, we also introduced the chemical potential $\mu$ that
should take the value $1/2$ for
standard Neveu-Schwarz boundary conditions on the circle, and the normal
ordering, $:(\ \ ):$, for regularizing the infinite Dirac sea. This is defined
by the ground-state conditions,
\be
  \begin{array}{l}
    c_k|\W\rangle =0, \quad k>0,\\ c_k^\dag |\W\rangle=0, \quad k\le 0,
  \end{array}
\qquad
:c^\dag_k c_k : = \left\{
  \begin{array}{l}
    \ \ c^\dag_k c_k,  \quad k>0,\\ -c_k c_k^\dag,  \quad k\le 0 .
 \end{array}
\right.
\label{no-ord}\ee

Note that the Laplace-transformed density (\ref{lap-def}) is by definition
the generating function of $\winf$ operators (\ref{lnm-def}) in the bulk;
after regularization, it accounts for the edge
operators,
\be
\r_k(\l)=\r_k -\l L_k +\frac{\l^2}{2} V^{(3)}_k +\cdots ,
\label{rho-ex}\ee
as defined in (\ref{lnm-edge}). The $R\to\infty$ subtractions (polynomial
in $R^2=N-1$) are rather remarkably accounted for.
For example, the ground-state values \cite{cdtz},
\be
\r_0|\W\rangle =\left(\frac{1}{2}-\mu\right)|\W\rangle, \qquad\quad
L_0|\W\rangle =\frac{1}{2}\left(\frac{1}{2}-\mu\right)^2|\W\rangle,
\qquad\left(\mu\to \frac{ 1}{2}\right),
\label{vac-val}
\ee
should be contrasted with the bulk values (\ref{q-m})
of order $N$ and $N^2$, respectively.
Therefore, the Laplace transformed density is the convenient
 quantity to establish the bulk-boundary correspondence.

We can now compute the algebra obeyed by the regularized
generators (\ref{rho-lim}) by using Fock space expressions.
The result is:
\be
\left[\r_k(\l),\r_n(\mu) \right]=
\left(t^{-\frac{n}{2}} y^{\frac{k}{2}}- t^{\frac{n}{2}}
  y^{-\frac{k}{2}} \right)\r_{k+n}(\eta)+c \; \d_{k+n,0}
\frac{z^{\frac{k}{2}}-z^{-\frac{k}{2}} }{z^{\frac{1}{2}}-z^{-\frac{1}{2}}},
\label{rho-alg}
\ee
where
\be
t=\frac{1}{1+\l}, \qquad y=\frac{1}{1+\d},\qquad z=ty=\frac{1}{1+\eta} .
\label{rho-par}
\ee
This is again the $\winf$ algebra (\ref{Winfty}) written in a basis that makes
it clear the map between the bulk and the edge: it is expressed in terms of
bulk operators but it realizes explicitly the conformal theory
representation. 
There appears the central extension proportional to $c\;\d_{k+n,0}$,
$c=1$, that originates from normal ordering.
Upon expanding (\ref{rho-alg}) in series of $\l$ and $\d$, and comparing with
(\ref{rho-ex}) one recovers the standard current algebra relations
(\ref{curr-alg}).
The highest-weight conditions are:
\be
\r_k(\l)|\W\rangle =0, \qquad \quad k\ge 0,\qquad \forall \l,
\label{hws}
\ee
because they hold for any current in the expansion (\ref{rho-ex}).

\subsection{Radial shape of neutral excitations}

A particle-hole excitation is described by the state
$|\{k\}\rangle=\r_{-k}|\W\rangle$, with momentum $k>0$.
The density fluctuation can be obtained from
the expectation value,
\be
\langle \r_0(\l)\rangle_{\{k\}}=
\frac{\langle \{k\}|\r_0(\l)|\{k\}\rangle}
{\langle \{k\}|\{k\}\rangle} =
\frac{\langle \W|\left[\r_k(0),[\r_0(\l),\r_{-k}(0)]\right]|\W\rangle}
{\langle \W|\left[\r_k(0),\r_{-k}(0)\right]|\W\rangle},
\qquad k>0,
\label{rho-ph}
\ee
that determines the radial profile by inverse Laplace transform.

Expression (\ref{rho-ph}) is computed by using the algebra (\ref{rho-alg}).
After restating the prefactor in (\ref{rho-lim1}), one finds the result:
\be
\langle \r_0(\l)\rangle_{\{k\}}=-
\frac{\left(1-(1+\l)^k \right)^2}{k\l (1+\l)^{R^2+k+1}} .
\label{rho-ph2}
\ee
This quantity only contains poles, since $R^2=N-1$, and its inverse
Laplace transform is computed by taking a path parallel to the imaginary
$\l$ axis and located to the right of all singularities, as usual.
Details of the calculation are given in the Appendix.
One finds the following expression in the edge limit (\ref{edge-lim2}):
\be
\langle \d\r(r)\rangle=\frac{1}{\pi}
\langle \d\r_0(r^2)\rangle_{\{k\}}=
\frac{4k x}{R^2}\frac{e^{-2x^2}}{(2\pi)^{3/2}},
\qquad r=R+x, \quad |x|<1 .
\label{rho-neu}
\ee

Of course, this result could have been directly obtained by using
fermionic Fock space. The expectation value over particle-hole
fluctuations reads:
\be
\langle \d\r(r)\rangle_{\{k\}}= \sum_{m'}
\left\vert \phi_{R^2+m'}\right\vert^2
\langle :c^\dag_{m'}c_{m'}:\rangle_{\{k\}}=\frac{1}{k}\left(
\sum_{i=1}^k \left\vert\phi_{R^2+i}\right\vert^2 -
\left\vert \phi_{R^2+i-k}\right\vert^2\right) .
\label{rho-ferm}
\ee
This quantity matches (\ref{rho-neu}) upon using the
expression of wavefunctions (\ref{field-op}) in the limit to the edge
\cite{cm},
\be
\phi_{R^2+m'}(r,\th)\sim\left(\frac{2}{\pi}\right)^{1/4}
\frac{e^{i(R^2+m')\th}}{\sqrt{2\pi R}}
  e^{-\left(x-\frac{m'}{2R} \right)^2} .
\label{wf-lim}
  \ee
The comparison provides a check for the previous analysis of the edge limit.
The advantage of the $\winf$ bosonic approach is that it can be
straightforwardly extended to fractional Hall states, as described
in Section 3.4.

\begin{figure}[h]
\begin{center}
  \includegraphics[width=0.8\textwidth]{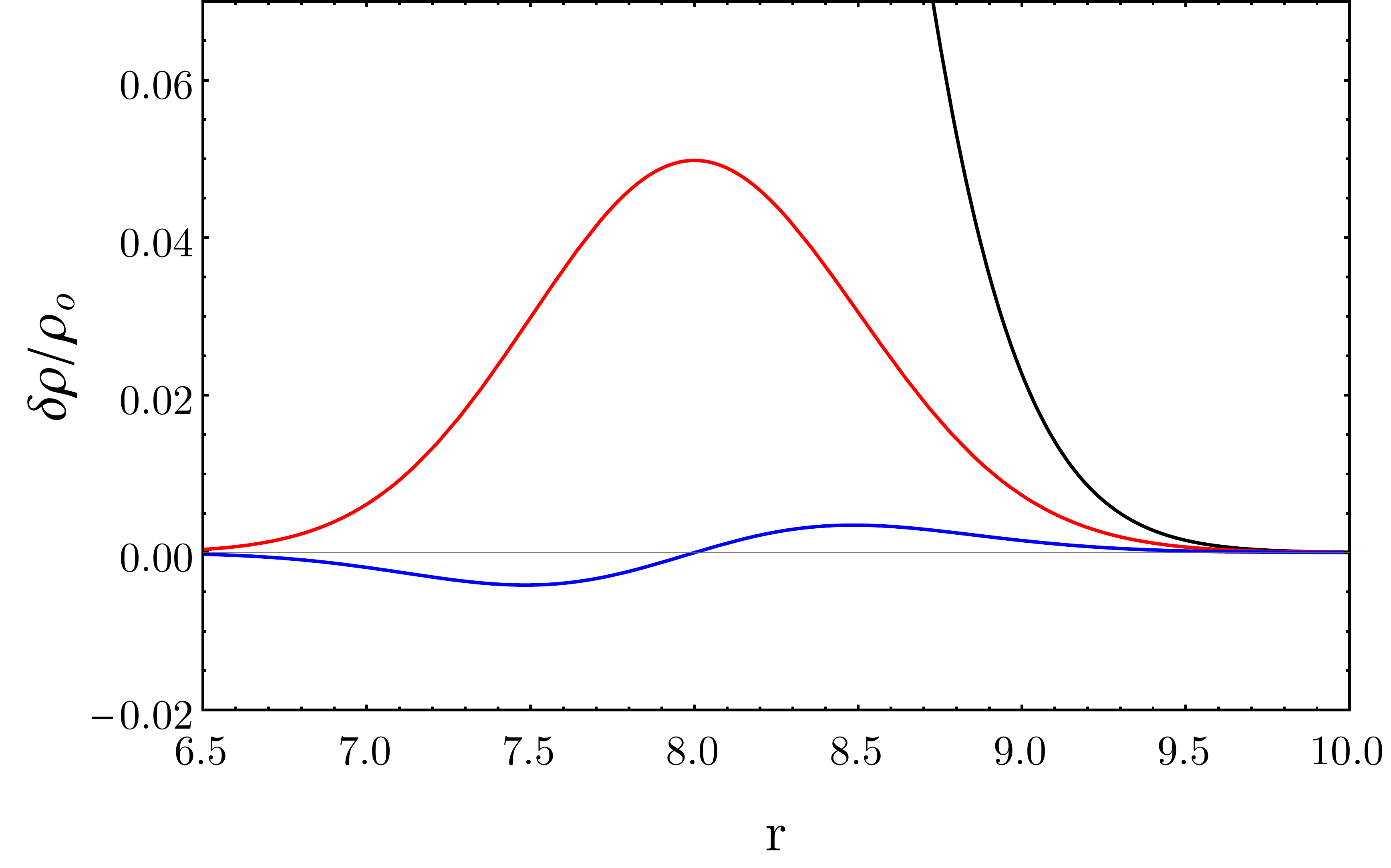}
  \caption{Radial density profile of $\nu=1$ ground state near the
    edge (normalized to one in the bulk, black line); $k=1$
    particle-hole excitation (blue); $Q=1$ charged edge excitation
    (red) (for $N=64$ electrons).}
\label{fig-edge}
\end{center}
\end{figure}

The density fluctuation (\ref{rho-neu}) is plotted in
Fig.\ref{fig-edge} together with the ground state profile near the
edge. One remarks the small size, $O(1/R^2)$ w.r.t. to the $O(1)$
ground state, and the support over $|x|=O(1)$.

\subsection{Shape of charged excitations}

The profile of a charged excitation at the edge can be obtained by following
similar steps. The corresponding state is $|Q\rangle = V_Q(0)|\W\rangle$,
where $V_Q(z)$ is the vertex operator of charge $Q$ in the $c=1$ conformal
theory.  The charged state is the highest-weight state of another
representation of the algebra and obeys the conditions,
\ba
&&\r_k(\l)|Q\rangle =0, \qquad\quad\ k>0, \qquad\forall \l ,
\label{vac-q}
\\
&& \r_0|Q\rangle =Q|Q\rangle, \qquad\quad L_0|Q\rangle =\frac{Q^2}{2}|Q\rangle,
\label{val-q}
\ea
analogous to those of the ground state  (\ref{vac-val}) and (\ref{hws}),
respectively.

The zero mode of the Laplace-transformed density, $\r_0(\l)$ is
actually the generating function of all higher-spin Casimirs,
whose eigenvalues are known from the study of
$\winf$ representations by the authors of Ref. \cite{kac}. 
We need to perform a change of parameters for interpreting their result:
the $\winf$ operators were presented in the form
$V\left(z^k \exp(a z\de_z)\right)$, being the generating
function of higher-spin operators (cf. (\ref{lnm-edge})).
Their value on charged states was found to be,
\be
V\left(z^k e^{a z\de_z}\right)|Q\rangle=
\frac{e^{aQ}-1}{e^a-1}|Q\rangle ,
\label{kac-val}
\ee
and the corresponding algebra was written,
\be
\left[V\left(z^k e^{a z\de_z}\right), V\left(z^n e^{b z\de_z}\right)\right]=
\left(e^{an}-e^{bk} \right)V\left(z^{k+n} e^{(a+b) z\de_z}\right)
+c\; \d_{k+n,0}\frac{e^{-ak}-e^{bk}}{1-e^{a+b}},
\label{kac-alg}
\ee
($c=1$ here).
These generators differ by our $\r_k(\l)$ in the ordering of derivatives.
The comparison of the two algebras, (\ref{rho-alg}) and (\ref{kac-alg}),
and the known data (\ref{val-q}) allows to find the relation between
the two parameterizations, with the result:
\be
V\left(z^{-k} e^{a z\de_z}\right)\equiv e^{a(k-1)/2}\r_k(\l),
\qquad \qquad e^a=\frac{1}{1+\l} .
\label{r-match}
\ee

The density profile for charged excitations is given by the expectation
value $\langle Q|\r_0(\l)|Q\rangle$. Upon using (\ref{kac-val})) with
the identifications (\ref{r-match}), we obtain:
\be
\langle \r_0(\l)\rangle_Q=
\frac{\langle Q|\r_0(\l)|Q\rangle}{\langle Q|Q\rangle}=
\frac{1-(1+\l)^{-Q}}{(1+\l)^{1/2}- (1+\l)^{-1/2}} .
\label{rho-q}
\ee
The inverse Laplace transform of
the expression (\ref{rho-q}) after inclusion of the bulk prefactor
$1/(1+\l)^{R^2+3/2}$,  leads to the result (see the Appendix):
\be
\langle \d\r(r)\rangle_Q =\frac{1}{\pi}\langle \r_0(r^2)\rangle_Q=
\frac{e^{-r^2}}{\pi}\sum_{n=N}^{N+Q-1}
\frac{r^{2n}}{\G(n+1)} .
\label{rho-qf}
\ee
In the $\nu=1$ case, this expression is rather simple, owing to
the integer values of $R^2=N-1$ and $Q$:
one recovers the sum of squared wavefunctions for $Q$ electrons added at the
edge. This result is actually exact, in spite of the 
approximations made in the limit to the edge.
Upon using the edge limit of wavefunctions (\ref{wf-lim}), we obtain the
expression:
\be
\langle \d\r(r)\rangle_Q =\left(\frac{2}{\pi}\right)^{1/2}
\frac{e^{-2x^2}}{2\pi R} Q+O\left(\frac{1}{R^2}\right),
  \qquad r=R+x, \quad |x|<1 .
\label{rho-exp}
  \ee
This fluctuation is definite positive, of size $O(1/R)$ and still
localized in a region of one magnetic length at the edge (see
Fig.\ref{fig-edge}).

In conclusion, we have shown that the study of 
the $\winf$ symmetry in the edge conformal field theory is able to
describe the radial shape of excitations, that is actually a two-dimensional
property of incompressible fluids. This kind of dimensional extension is
possible thanks to the existence of an infinite tower of
conserved currents.
In the discussion of Section two, we said that the radial dependence
is lost in the limit to the edge, but this is not actually true.


\subsection{Density profiles for Laughlin states}

The $\winf$ symmetry description of edge excitations extends to Laughlin
states with filling fractions $\nu=1/m=1/3,1/5,\dots$.
The algebra (\ref{rho-alg}) of $\r_k(\l)$ is unchanged but the representations
are different: the generating function of Casimirs (\ref{rho-q}) takes
the same form, but the charge $Q$ is no longer integer quantized.

As explained in
Section 2, the $\winf$ algebra with central charge $c=1$ contains the
current algebra (\ref{curr-alg}) of charge and angular momentum
operators, that has been extensively analyzed for Laughlin states
 \cite{cdtz}. Their eigenstates have been determined by e.g. canonical
quantization of the chiral boson theory, and read:
\be
\r_0|\frac{n}{m}\rangle=
\frac{n}{\sqrt{m}}|\frac{n}{m}\rangle,\qquad
L_0|\frac{n}{m}\rangle=
\frac{n^2}{2m}|\frac{n}{m}\rangle,\qquad\ \ n\in\Z,
\quad \left( \nu=\frac{1}{m} \right).
\label{bose-val}
\ee
Note that the physical charge is $Q=n/m$, but the eigenvalue
$\wt Q=n/\sqrt{m}$
of $\r_0$ differs by a proportionality factor due to our normalization
of the current algebra (\ref{curr-alg}).

As explained in the previous section, the Laplace
transformed density for charged excitations is obtained from
the generating function of Casimirs, Eq.(\ref{rho-q}).
After inclusion of the bulk prefactor, it reads:
\be
\langle\wt\r_0(\wt\l)\rangle_{\wt{Q}} = \frac{1-(1+\wt\l)^{-\wt Q}}
{\wt\l(1+\wt\l)^{\wt R^2+1}}
\sim \wt Q-\wt\l\left(\frac{\wt Q^2}{2}+
  \wt Q\left(\wt R^2+\frac{3}{2}\right)\right)+O(\wt \l^2) .
\label{rho-bose}
\ee
The first two terms in the expansion in $\wt\l$ 
specify the edge charge $\r_0$ and angular momentum $L_0$ eigenvalues.
Using the spectrum (\ref{bose-val}), we find the following result for
the excitation of $q$ electrons added at the boundary, 
\be
\langle\wt\r_0(\wt\l)\rangle_{\wt{Q}} \sim \sqrt{m}q-
\wt\l\left(\frac{mq^2}{2}+q\sqrt{m}\left(\wt R^2+\frac{3}{2}\right) \right)
+O(\wt \l^2) .
\label{rho-bose2}
\ee

This expression should be compared with the approximate knowledge of
bulk data for fractional fillings. The first two moments of the
density, in presence of $q$ added electrons, are:
\ba
&&\int d^2r \left(\r_{N+q}(r) -\r_N(r)\right)=q,
\nl
&&
\int d^2r\; r^2 \left(\r_{N+q}(r) -\r_N(r)\right)=\frac{m}{2}
\left(q^2+2q\left(N+O(1)\right)\right) .
\label{rho-sum}
\ea
where the term $O(1)$ as $N\to\infty$
depend on the form of the electron droplet near
the edge, that is not relevant in the present discussion
(The ground state profile will be analyzed in the next Section).
Therefore, we can infer the bulk expression:
\be
\langle \r_0(\l)\rangle_q\sim q-
\l\left(\frac{mq^2}{2}+q\left(mN+O(1)\right)\right) + O(\l^2),
\label{rho-bulk}
\ee
where $R^2=mN+O(1)$ for $N\to\infty$.
The comparison of edge (\ref{rho-bose2}) and bulk (\ref{rho-bulk}) results
leads to the following identification of parameters,
\be
\r_0(\l)=
\frac{\wt\r_0(\wt\l)}{\sqrt{m}},
\qquad \wt\l=\sqrt{m}\l, \qquad \wt R^2=\frac{R^2}{\sqrt{m}} .
\label{rho-match}
\ee
After the matching, the density of charged excitations is finally found
to be:
\be
\langle\r_0(\l)\rangle_Q =\frac{1}{m}\;
\frac{1-(1+\sqrt{m}\l)^{-\frac{n}{\sqrt{m}} }}
  {\l(1+\sqrt{m}\l)^{\frac{R^2}{\sqrt{m}} }},
  \qquad\quad Q=\frac{n}{m}, \quad \nu=\frac{1}{m} .
  \label{rho-frac}
\ee

The inverse Laplace transform determines the following radial profile
(see Appendix for the derivation),
\be
\langle \d \r(r)\rangle_Q=\frac{1}{m\pi}\left(
  \frac{\Gamma\left(\frac{R^2+n}{\sqrt{m}},\frac{r^2}{\sqrt{m}}\right)}
  {\Gamma\left(\frac{R^2+n}{\sqrt{m}}\right)} - (n=0)\right),
\qquad\quad Q=\frac{n}{m},
\label{rho-gam}
\ee
where $\Gamma(a,z)$ is the incomplete Gamma function.
The profile can be made more explicit by expanding near the edge, i.e.
$r=R+x$, $R\to\infty$ (see Appendix):
\be
\langle \d \r(r)\rangle_Q=
\frac{e^{-\frac{2x^2}{\sqrt{m}} }}{m^\frac{1}{4}\sqrt{2\pi}\pi R} Q,
\qquad\quad Q=\frac{n}{m}, \quad \nu=\frac{1}{m} .
\label{rho-qfr}
\ee
This expression is similar to the $\nu=1$ case (\ref{rho-exp}),
with the Gaussian
width enhanced by the factor $\sqrt{m}$; note, however, that it
cannot be realized with a finite number of fermion Fock space states,
as expected by bosonization on general grounds.

\begin{figure}[h]
\begin{center}
\includegraphics[width=0.8\textwidth]{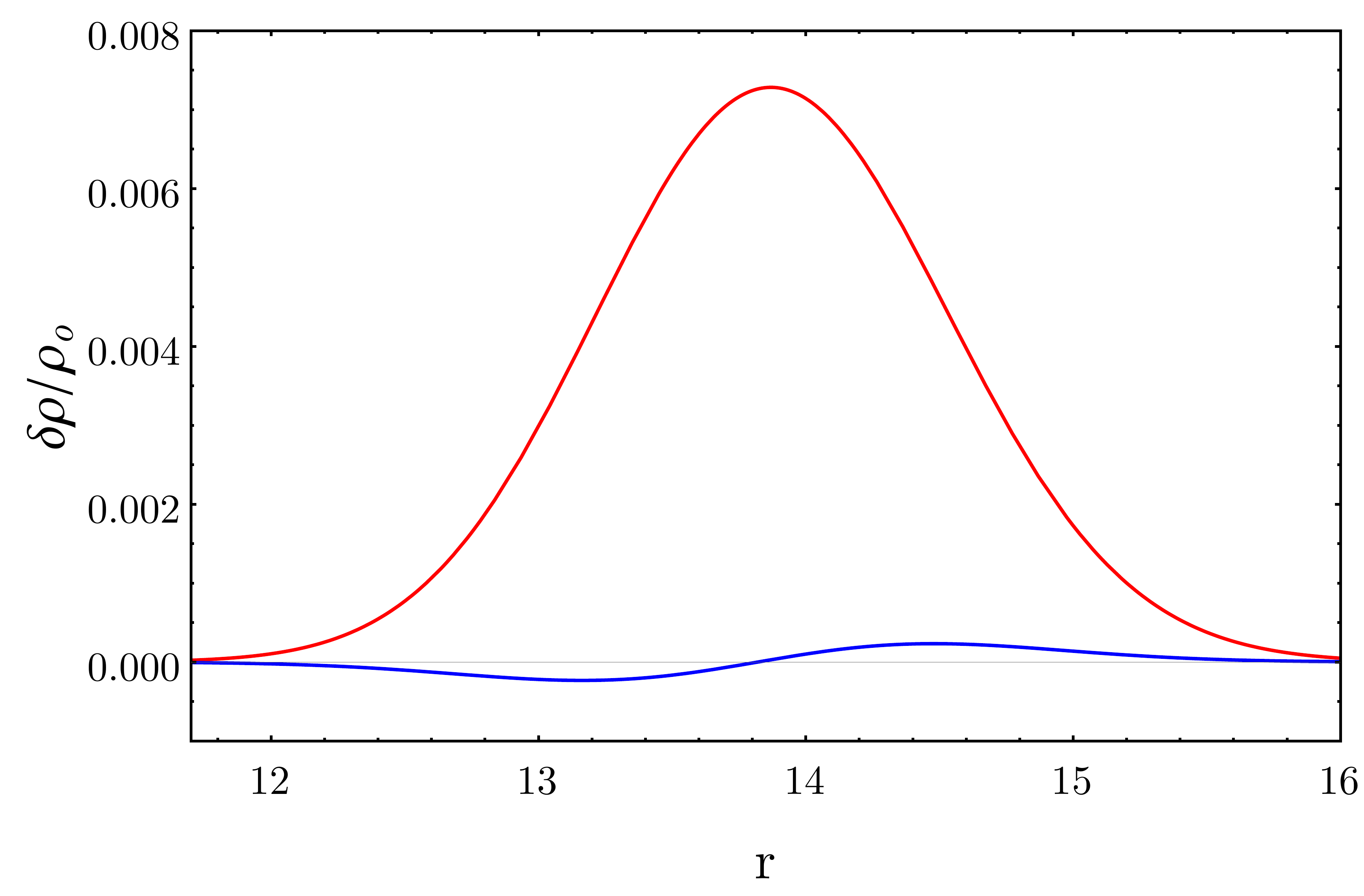}
\caption{Radial density profile of $k=1$ neutral edge excitation
  (blue) and $Q=1/3$ charged fluctuation (red) for $\nu=1/3$.  The
  ground-state density profile terminates at $r\sim 15$ (not plotted)
  ($N=64$).}
\label{fig-frac}
\end{center}
\end{figure}

The derivation of density profile for neutral particle-hole excitation
over the Laughlin state follows similar steps. Upon using the bulk-boundary
parameter matching (\ref{rho-match}), we obtain:
\be
\langle \d\r(r)\rangle_{\{k\}}=
\frac{4k x}{R^2}\frac{e^{-\frac{2x^2}{\sqrt{m}} }}
  {(2\pi)^{\frac{3}{2}}m^{\frac{3}{4}}   }, \quad
  \qquad r=R+x, \quad |x|<1 .
  \label{rho-net}
\ee
These results are shown in Fig.\ref{fig-frac} for $\nu=1/3$:
a particle-hole excitation, the $Q=1/3$ charged excitation and
the ground state profile for $N=64$ obtained numerically in
Ref.\cite{laugh-num}.

We conclude this section with some remarks:
\begin{itemize}
\item As anticipated, the bosonic approach given by $\winf$ symmetry
  extends to fractional states: well-known conformal field theory
  results on states and quantum numbers of edge excitations are
  complemented here by the density profiles.
\item
  Being relativistic, this approach is based on normal-ordered
   quantities and cannot predict the density profile of the ground
   state, that is rather non-trivial for fractional fillings, as
   discussed in the next Chapter.
 \item
   The parameter matching (\ref{rho-match}) for fractional fillings
     involves rescaling the density $\r_0(\l)$ and the magnetic length
     $\ell^2\to \sqrt{m}\ell^2$. This is different from the naive
     scaling $\ell^2\to m\ell^2$ suggested by the composite fermion
     map \cite{jain}.  A simple, intuitive argument for this result is
     presently missing.
   \item
     The prediction for the density profile (\ref{rho-gam}) can be
     compared with the numerical results of Ref. \cite{abanov} by
     considering densities of nearby electron numbers. The analytic
     studies of the density profile in Ref.  \cite{edge-bump} could
     also provide a check, but require the extension of our approach
     to the cylinder geometry.
\end{itemize}

\subsection{Universality of density profiles}

The previous derivation of the density fluctuations is based on algebraic data
of the conformal field theory and thus possess universal features. Let us
discuss this point in more detail.

Our assumption are:
\begin{itemize}
\item
  Conformal invariance of massless edge modes
as described by the $c=1$ bosonic theory (chiral
Luttinger theory). The existence of higher-spin conserved currents
is automatic in this theory.
\item
  The matching with bulk physics is implemented by the geometrical
  picture of the quantum incompressible fluid, that is proven for
  $\nu=1$ and assumed for $\nu=1/m$, following Laughlin. This implies
  the $W_\infty$ symmetry and the correspondence between infinite
  conserved currents (edge) and radial moments of the density (bulk).
  The bulk introduces the magnetic length as unique scale in the
  description.
\item
  No specification of the Hamiltonian is needed, besides the
  assumption of a gap for density fluctuations and short-range
  residual interactions, underlying   the incompressible fluid picture.
\end{itemize}

We conclude that density fluctuations (\ref{rho-qfr}) and (\ref{rho-net})
are `geometrical' data of
Hall states, meaning that they are robust under deformations
of the microscopic dynamics that keep the gap open and do not spoil
the incompressible fluid. Besides, a short-range Hamiltonian can be
included in our setting, as described in the next Section.


\section{$\winf$ symmetry of bulk excitations}

\subsection{Boundary layer of Laughlin states}

\begin{figure}[h]
\begin{center}
\includegraphics[width=1\textwidth]{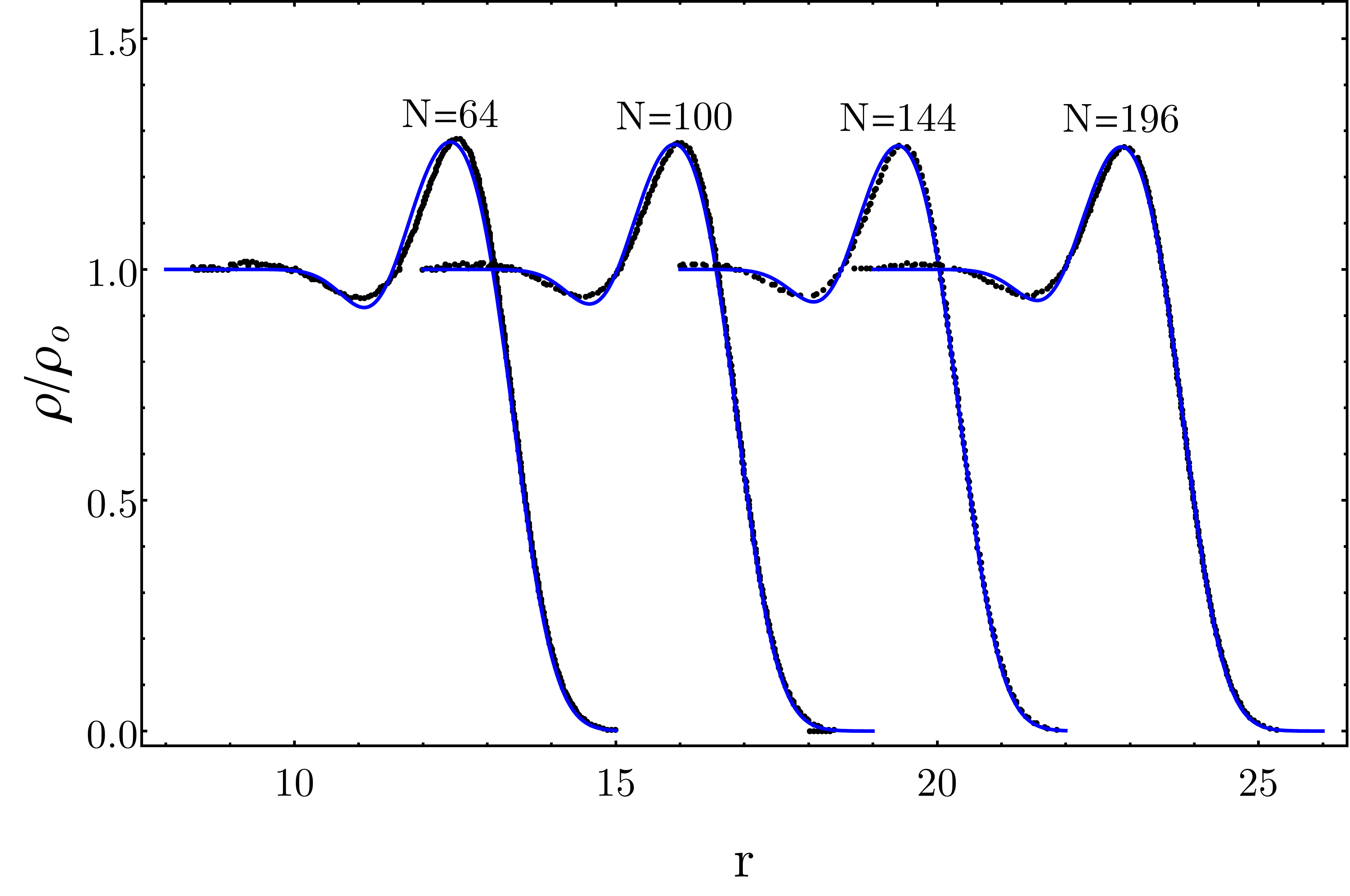}
\caption{Numerical density profile of the $\nu=1/3$ Laughlin ground state for
$N=64,100,144, 196$ \cite{laugh-num} (black points) and two-block fit (blue).}
\label{fig-shoot}
\end{center}
\end{figure}

In this section we analyze the density profile of Laughlin states as an
introduction to the following discussion of bulk excitations.  The numerical
evaluation of the density for $\nu=1/3$ is shown in Fig.\ref{fig-shoot} for
several values of electron number $N$ \cite{laugh-num}.  One sees the presence
of a boundary layer or `overshoot', whose size is a few magnetic lengths and
approximately $N$-independent.

The overshoot is the result of highly non-trivial correlations inside
the Laughlin state, but for what concern the density profile, it is possible
to model it in terms of fermionic occupation
numbers $n_j$ for momentum $j$, as follows:
\be
\r_o(r)=\langle\W| \r(z,\bar z)|\W\rangle_{\nu=1/3}= \sum_{j=0}^\infty
\vert \phi_j(z,\bar z)\vert^2 \ n_j .
\label{rho-fock}
\ee
The $n_j$ correspond to the average over occupation numbers of
very many Slater determinants contained in the Laughlin state.

\begin{figure}[h]
\begin{center}
\includegraphics[width=0.6\textwidth]{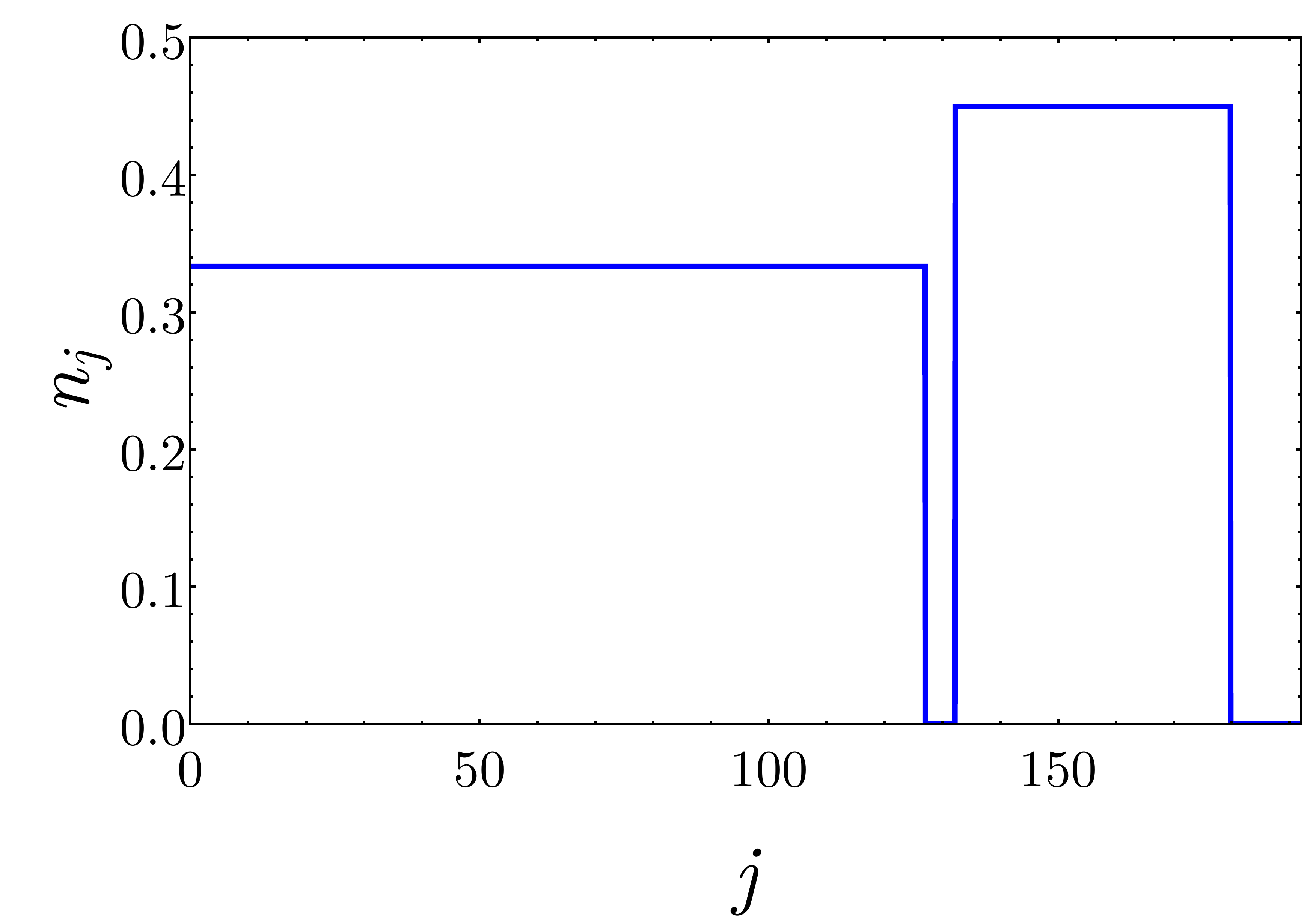}
\caption{Occupation numbers $n_j$ of single particle states with momentum
$j$ for fitting the $\nu=1/3$ Laughlin ground state density ($N=64$).}
\label{fig-block}
\end{center}
\end{figure}

Let us discuss a simple two-block distribution of occupation numbers,
as show in Fig.\ref{fig-block}.
The first block is given by the flat droplet with $n_j=1/3$
but shortened by $\D j= -a$ from its natural extension
$j_{\rm max}=3N-1$. The second block of variable height $n_j=h$
extends for $-b <\D j\le -c$. These occupation numbers can actually be realized
by a finite sum of Slater determinants.
Two of the four parameters $\{a,b,c,h\}$ are fixed by the sum rules
obeyed by the Laughlin state:
\be
\int d^2r\ \r_o =N,\qquad \int d^2r\; r^2\r_o =\frac{m N(N-1)}{2}+N,
\qquad \nu=\frac{1}{m}
\label{rho-sum2}\ee
while the two others are approximatively determined by the fit.
The result are:
\be
a\sim 8.00 \sqrt{N},\qquad
b\sim 7.35 \sqrt{N},\qquad
c\sim 1.40 \sqrt{N},\qquad
h\sim 0.45,
\label{ov-fit}\ee
for $N=64,100,144,196$, with error of about $3$ on the last digit.
Further analysis of the $N=200$ data of Ref.\cite{abanov} confirms the
values (\ref{ov-fit}).
The results of the fit are shown in Fig.\ref{fig-shoot}.

This analysis shows that the overshoot is characterized by
angular momentum values $\D j\sim\sqrt N$. Since $R^2\sim m N$,
its radial extension, $\D r\sim m\D j/(2R)$, approaches
a constant for $R\to\infty$.
Furthermore, the size $\D \r$ of the modulation is also independent
of the number of electrons (see Fig.\ref{fig-shoot}).

It is apparent that the edge
excitations discussed in the previous Section are much smaller than the
overshoot: they correspond to small
deformations of the density outer slope and are manifestly unrelated
to the non-trivial profile inside the droplet. Let us compare
the respective sizes:
\be 
\begin{array}{l|l|l|l}  
  & \D j  & \D r  & \D \r  \\
  \hline
{\rm overshoot}  &O( R) & > 1 & O( 1) \\
  {\rm edge\ excitation} &O(1)  & O(1)& \le \frac{1}{R} \\
\end{array}
\label{ex-comp}
\ee

The shape of the overshoot cannot be determined in our relativistic
approach based on normal-ordered quantities, as already said.
Nevertheless, its properties suggest us to extend
the $\winf$ symmetry analysis
to the regime of `large edge excitations', possessing a
finite $R\to\infty$ limit in terms of size, momenta and energies.
In the following, we shall find that this generalization 
can be done and will give access to analytic results for bulk physics.

Let us add a side remark. The ground state density profile for other
filling fractions $\nu=1/5,1/7,\dots$ presents more that one
oscillation at the boundary and therefore the two-block fit must be
replaced by a multi-block form. As explained in Ref.\cite{abanov}, these
fluctuations near the boundary correspond to the preemptive freezing
of Hall electrons into a Wigner crystal, starting from the outer parts
of the droplet.


\subsection{$\winf$ algebra in the bulk regime}

\subsubsection{The $R\to\infty$ limit for large excitations}
  
We shall now extend our analysis to large edge excitations that
possess a finite $R\to\infty$ limit.
We reconsider the earlier derivation of the $\winf$ algebra for $\nu=1$
in Section 3.1, but perform another $R\to\infty$ limit, characterized by
the following ranges of coordinates and angular momenta,
\be
r=R+x, \quad x=O(1),\qquad m=R^2+ m',\quad m'=O(R),
\qquad {\bf k} =\frac{m'}{R}=O(1),
\label{large-exc}\ee
where ${\bf k}$ is the momentum of waves propagating along the
edge, approximately straight for $R\to\infty$.

Upon taking this limit in the expression of the Laplace transformed
density (\ref{lap-alg}), one finds the same expression up to an exponential
prefactor (see the Appendix for the derivation),
\be
\r_n(\l) \quad\rightarrow\quad e^{-\frac{n^2}{8R^2}}\r_n(\l) ,
\label{large-rho}\ee
and their algebra (\ref{rho-alg}) is modified as follows:
\be
\left[\r_k(\l),\r_n(\mu) \right]=
 A_{k,n}(\l,\mu)\;\r_{k+n}(\eta)+\d_{k+n,0}\;B_k(\eta) ,
\label{large-alg}\ee
where
\ba
&& A_{k,n}(\l,\mu)=e^{\frac{kn}{4R^2}}
\left(x^{-\frac{n}{2}} y^{\frac{k}{2}} -
x^{\frac{n}{2}} y^{-\frac{k}{2}}\right),
\nl
&&B_k(\eta)=e^{-\frac{k^2}{4R^2}}\sum_{i=1}^k z^{i-\frac{k+1}{2}},
\qquad k>0 ,
\nl
&& t=\frac{1}{1+\l}, \quad y=\frac{1}{1+\d},
\quad z =t y=\frac{1}{1+\eta} .
\label{large-val}\ea

Note that the $\winf$ algebra is basically unchanged w.r.t. (\ref{rho-alg}).


\subsubsection{From Laplace to Fourier modes}

Another useful observation about the $R\to\infty$ limit is that
the radial coordinate $x=r-R$ actually become unbounded from below
for density fluctuations that have finite support.
Therefore, the Laplace transform on $r^2>0$ can be replaced
by the Fourier transform in $x\in \RR$.
The relation between parameters are easily determined,
\be
\l=-\frac{\de}{\de r^2}= -\frac{1}{2r}\frac{\de}{\de r}
\sim -\frac{1}{2R}\frac{\de}{\de x}=-\frac{i{\bf p}}{2R}, \qquad\quad
R\to\infty,
\label{f-def}\ee
where ${\bf p}=O(1)$ is the momentum conjugate to $x=O(1)$ and thus orthogonal
to the boundary.

Laplace and Fourier transforms of the density are identified as follows:
\be
\r_k(\l)\ \rightarrow\ \r_k({\bf p}), \qquad\quad
\frac{1}{1+\l}\sim \exp\left(i\frac{{\bf p}}{2R}\right),
\label{f-def2}\ee
and the $\winf$ algebra becomes,
\be
\left[\r_k({\bf p}),\r_n({\bf p'}) \right]=
A_{k,n}({\bf p},{\bf p'})\;\r_{k+n}({\bf p+p'})+
\d_{k+n,0}\;B_k({\bf p+p'}),
\label{f-alg}\ee
where
\ba
&& A_{k,n}({\bf p},{\bf p'})=
e^{\frac{kn}{4R^2}}\left(
  e^{\frac{i}{4R}(k{\bf p'}-n{\bf p})}-
  e^{\frac{-i}{4R}(k{\bf p'}-n{\bf p})}\right),
\nl
&&B_k({\bf p+p'})=e^{-\frac{k^2}{4R^2}}
\sum_{\ell=1}^k e^{i\frac{\bf p+p'}{2R}\left(\ell-\frac{k+1}{2}\right)},
\qquad k>0 .
\label{f-val}\ea
Note that the $1/R$ dependence should be kept, because
the indeces are $k,n=O(R)$: a finite $R\to\infty$ limit is
achieved at the end of calculations.

The algebra (\ref{f-alg}) is similar to the Girvin-MacDonald-Platzman
form (\ref{gmp-alg}), but there are two crucial differences:
\begin{itemize}
\item
  There is a central extension that originates from relativistic
  normal ordering and the norm of the Hilbert space for the edge theory.
\item Ground state conditions as well as representations
  (values of Casimirs) are known. This permits analytic computation of
  observables as discussed in the following sections.
    \end{itemize}

\subsection{Bosonization of the short-range interaction}

We now find the form of the Hamiltonian in the regime of large fluctuations.
Consider the Gaussian two-body potential, parameterized by the variable
$t>0$:
\be
H_t=\frac{1}{2}\int d^2 r_1 d^2 r_2\; \r(\vec{r}_1)\;
\exp\left(-t|\vec{r}_1-\vec{r}_2|^2\right)\; \r(\vec{r}_2) .
\label{h-gauss}\ee
This type of short-range potential yields relatively simple
matrix elements for $R\to\infty$. Furthermore, by expanding in
the $t\to\infty$ limit, one obtains the Haldane ultra-local potentials
\cite{gp}. For example:
\ba
\d^{(2)}(\vec{r}_1-\vec{r}_2)&=&\lim_{t\to\infty}
\frac{t}{\pi}e^{-t|\vec{r}_1-\vec{r}_2|^2},
\label{del-p}\\
\nabla^2\d^{(2)}(\vec{r}_1-\vec{r}_2)&=&\lim_{t\to\infty}
\frac{-t^2}{\pi}\left(1+t\frac{\de}{\de t}\right)
e^{-t|\vec{r}_1-\vec{r}_2|^2} .
\label{del-h}\ea
The first interaction actually vanishes for fermionic systems, while
the second one is the first Haldane potential for which the $\nu=1/3$
Laughlin state is the exact ground state.

The evaluation of the two-body matrix elements in the $\nu=1$ fermionic Fock
space and their $R\to\infty$ limit (\ref{large-exc}) by saddle-point
approximation is carried out in the Appendix, by extending results of
Ref.\cite{ctz-dyna}.  After antisymmetrization with respect to fermion
exchanges, the result is indeed found to behave as $O(1/t^2)$ for large $t$,
thus checking the vanishing of the delta interaction (\ref{del-p}).  This
leading behavior determines the Haldane potential we are interested in. We
obtain:
\be
H= -\frac{2}{R}\sum_{r,s,k\in \Z} e^{-\frac{k^2}{4R^2} -\frac{(r-s+k)^2}{4R^2}}
\left(\frac{k^2}{4R^2} -\frac{(r-s+k)^2}{4R^2}\right) c^\dag_s c_{s-k}
c^\dag_r c_{r+k} .
\label{h-hald}\ee
In this expression, we omitted inessential numerical factors and
used angular momentum indeces $\{r,s,k\}$ shifted to the edge, as in
(\ref{large-exc}). 
The quadratic forms appearing in the matrix element
are actually dictated by antisymmetrization.

The Hamiltonian can be bosonized using the following trick.
Note that the first term in the exponential in (\ref{h-hald}) is the prefactor
for $\r_k(\l)$ in (\ref{large-rho}). The second term is actually the
square of the coefficient in its Fock space expression 
(cf. (\ref{rho-lim}) and (\ref{f-def2})):
\be
\r_k({\bf p})=e^{-\frac{k^2}{8R^2}}\sum_r
  e^{i\frac{\bf p}{2R}\left(r+\frac{k-1}{2}\right)} c^\dag_r c_{r+k} .
\label{rho-f}\ee
Using the Gaussian integral we can write,
\ba
H &=& -
\frac{2}{R}\sum_{r,s,k\in \Z}
\int_{-\infty}^{\infty} \frac{d {\bf q}}{2\sqrt{\pi}}
e^{-\frac{{\bf q}^2}{4}}
\left(\frac{k^2}{4R^2}+\left(\frac{\de}{\de {\bf q}}\right)^2\right)
e^{i\frac{{\bf q}}{2R}\left(r+\frac{k}{2} -s +\frac{k}{2}\right)}
e^{-\frac{k^2}{4R^2}}
c^\dag_s c_{s-k} c^\dag_r c_{r+k}
\nl
&=&- \frac{1}{R}\sum_{k > 0}
\int_{-\infty}^{\infty}  \frac{d {\bf q}}{2\sqrt{\pi}}  \
e^{-\frac{{\bf q}^2}{4}}
\left(\frac{k^2}{R^2}+{\bf q}^2-2\right)
\r_{-k}(-{\bf q}) \r_k({\bf q}) .
\label{h-bose}\ea
In this expression, normal ordering with respect to the ground state
(\ref{no-ord}) has been implemented, leading to $k \ge 0$.
The  $k=0$ term is singled out with factor one half \cite{cdtz}:
however, this is non-vanishing on charged states (\ref{val-q}) only,
and should be removed for later evaluation of excitation spectra.

In conclusion, we have shown that the short-range interaction can be bosonized
in the regime of large fluctuations. The resulting expression looks similar to
the Fourier transform of the Haldane potential (\ref{del-h}), in terms of
momenta $\{{\bf q}, {\bf k}=k/R\}$, orthogonal and longitudinal to the edge,
respectively. Note the condition $k> 0$ due to chirality.

Let us add some remarks:
\begin{itemize}
  \item
In the case of edge excitations, the momenta $\{r,s,k\}$ take
finite values: approximating the exponential factor in (\ref{h-hald})
by one, we would get the Hamiltonian $H\sim\sum(k/R)^2 \r_{-k}\r_k$
and recover the spectrum of capillary waves,
$\eps_k\sim - k^3/R^3$, as discussed in Ref. \cite{ctz-dyna}.
In the present case, we are interested in large fluctuations $k=O(R)$, thus
the exponentials should be kept.
\item
The form of the Hamiltonian (\ref{h-bose}) can be modified by adding a
one-body term: this can be done e.g. for enforcing translation
invariance of the electron droplet, requiring a vanishing energy
for ${\bf q}=0$ and $k=0,\pm 1$ \cite{ctz-dyna}.
However, this is not relevant for the following discussion of
the $k=O(R)$ regime.
\end{itemize}

\subsection{Spectrum of large neutral excitations:
  edge reconstruction}

We reconsider the basic particle-hole edge excitation of Section 3.2,
described by:
\be
|\{{\bf n, p}\}\rangle=\r_{-n}({\bf p})|\W\rangle ,
\label{n-ex}\ee
where ${\bf n}=n/R >0$ is the momentum parallel to the edge
in the extended range (\ref{large-exc}). We also allow
a non-vanishing orthogonal momentum ${\bf p}$.

The energy of this excitation is given by the
expectation value:
\be
  \eps({\bf n,p})=
  \frac{\langle\{{\bf n, p}\}|H |\{{\bf n, p}\}\rangle}
  {\langle\{{\bf n, p}\} |\{{\bf n, p}\}\rangle} .
\label{n-def}\ee
The expression to be evaluated at the numerator is,
\be
- \frac{2}{R\sqrt{\pi}}\sum_{k> 0}
\int_{-\infty}^{\infty}  d {\bf q}\; e^{-\frac{{\bf q}^2}{4}}
\left( \frac{k^2}{4R^2}+ \frac{\de^2}{\de^2{\bf q}}\right)
\langle \W| \r_{n}({\bf -p})\r_{-k}({\bf -q}) \r_{k}({\bf q}) 
  \r_{-n}({\bf p})|\W\rangle .
\ee
Using the ground state conditions (\ref{vac-q}),
\be
\r_n({\bf p})|\W\rangle=0,\qquad\quad n \ge 0,\quad \forall\; {\bf p} ,
\label{hws-f}\ee
the result is obtained by repeatedly using the commutators (\ref{f-alg}).
Note that the bulk prefactor (\ref{rho-lim1}) for the density is not needed,
since it cancels out in the expression (\ref{n-def}).
The result is (see the Appendix for details):
\ba
\!\!\! \! \! \eps({\bf n,p}) &=&-\frac{2}{\bf n} 
\int_0^{\bf n} d {\bf k}\; e^{-\frac{{\bf k}^2}{4}} \left({\bf n-k}\right)
\left[{\bf k}^2+\left({\bf n}^2-{\bf k}^2\right)
  \cos\left(\frac{\bf p k}{2} \right)e^{-\frac{{\bf n}^2}{4}}\right]
\nl
& -&  \frac{e^{-\frac{{\bf n}^2}{4}}}{\bf n}
\int_0^{\bf n} d  {\bf k}\int_0^{\bf n} d  {\bf k'}
e^{-\frac{({\bf k - k'})^2}{4}}
\left({\bf n}^2-({\bf k - k'})^2\right)
  \cos\left(\frac{\bf p (k-k')}{2} \right).
\label{n-spec}\ea
Note that this expression has a finite $R\to\infty$ limit once expressed
in terms of momenta ${\bf n}$ and ${\bf p}$, as promised.

\begin{figure}[h]
\begin{center}
\includegraphics[width=0.9\textwidth]{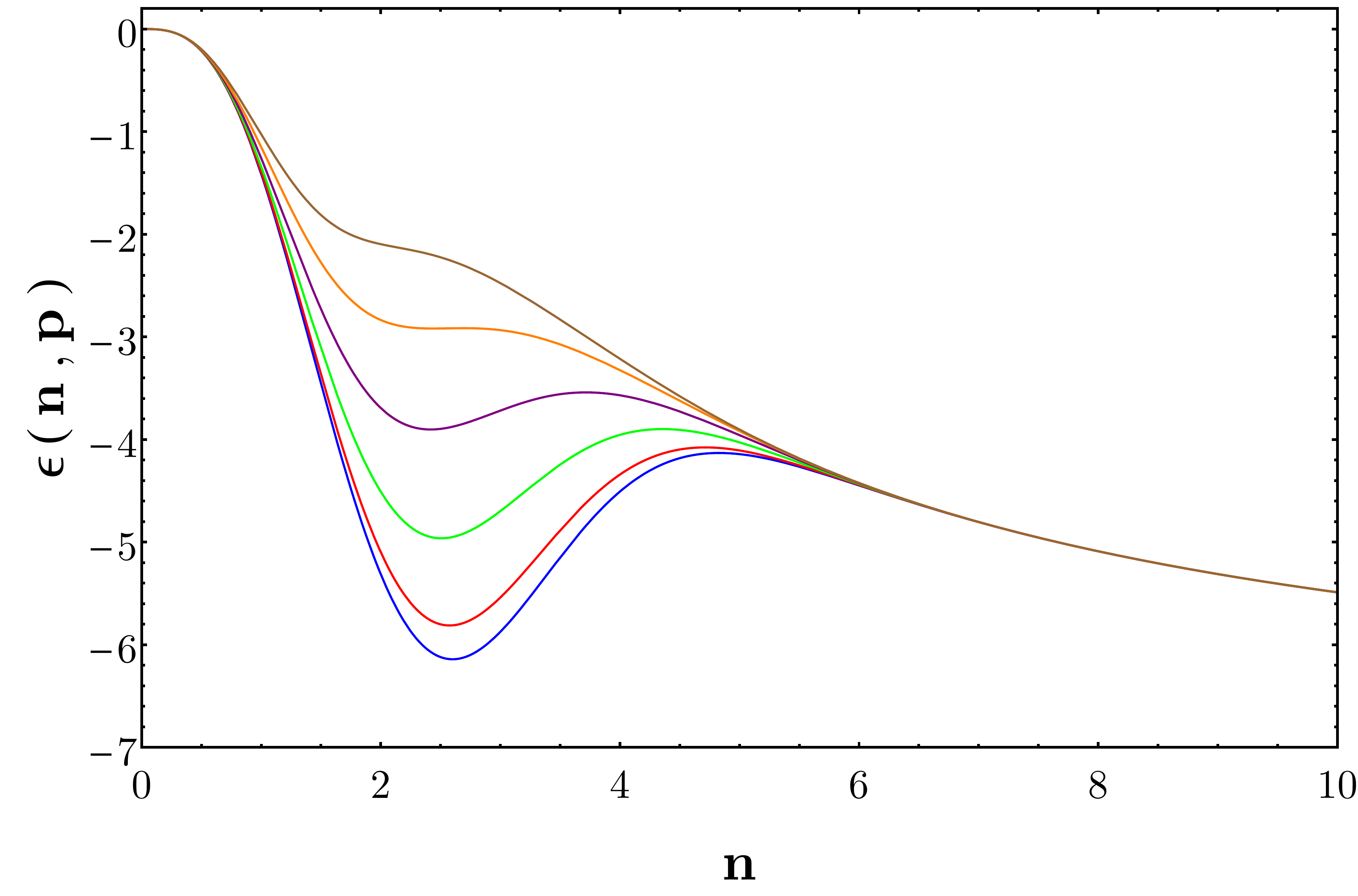}
\caption{Energy spectrum $\eps ({\bf n,p})$ of $\nu=1$ large neutral
  excitations plotted as a function of ${\bf n}$ for values of
  ${\bf p }=0,1,\dots,5 $ (from bottom to top).}
\label{fig-spec}
\end{center}
\end{figure}

\begin{figure}[h]
\begin{center}
\includegraphics[width=0.9\textwidth]{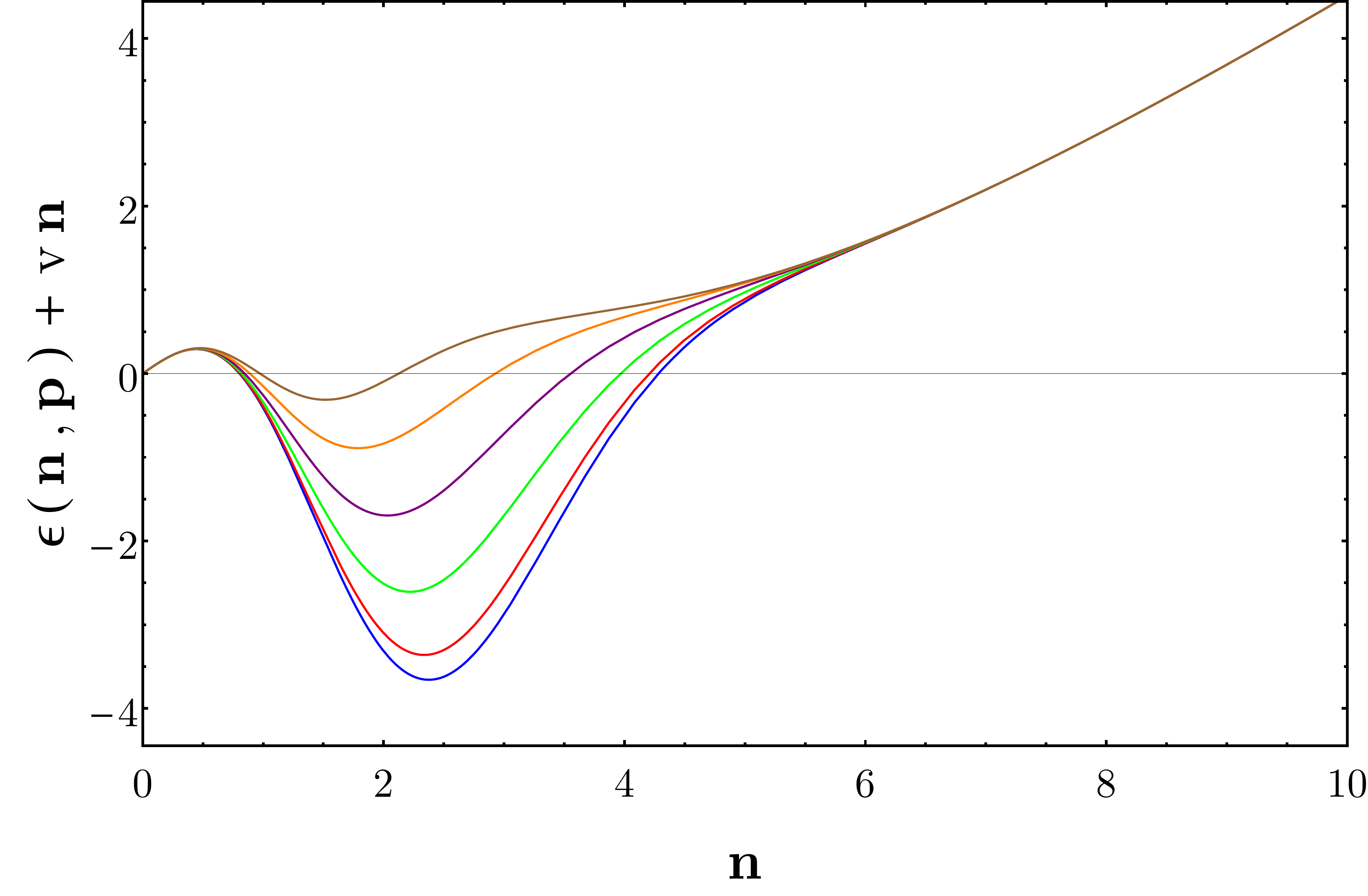}
\caption{Energy spectrum $\eps ({\bf n,p})+v{\bf n}$ of $\nu=1$ large
  neutral excitations plotted as a function of ${\bf n}$ for values of
  ${\bf p }=0,1,\dots, 5 $ (from bottom to top).}
\label{fig-rot}
\end{center}
\end{figure}

The resulting energy spectrum is plotted in Fig.\ref{fig-spec} as a
function of the momentum ${\bf n}$ for several values
of  ${\bf p}$. One can distinguish three behaviors:
\begin{itemize}
\item
  the edge regime for small ${\bf n}$, corresponding to the capillary waves,
  $\eps({\bf n,p})\sim - {\bf n}^3$ \cite{ctz-dyna}.
\item
  A local minimum for ${\bf k}={\bf k}_o\sim 2.6$ and ${\bf p}\lesssim 10$.
\item
  A decreasing behavior asymptotically reaching a constant for large
  $\bf n$ (not seen in Figure).
  \end{itemize}
Note that the Haldane potential is classically repulsive and this explains the
decreasing and saturation of the spectrum at large {\bf k}.
There is however a fermionic
exchange term that is attractive and causes the local minimum. 
This phenomenon is the so-called edge reconstruction,
the tendency of the incompressible fluid droplet to split a thin ring
to finite distance $x_o=O(\ell)$ \cite{edge-reco}. 
Originally observed for  Coulomb interaction, 
it is also present for short-range potentials.
Upon adding a standard quadratic confining potential, corresponding to
$\eps({\bf n,p})\to \eps({\bf n,p}) +v {\bf n}$, one finds
a minimum at finite ${\bf k} \lesssim 2.6$,
called edge roton \cite{jain-rot} (see Fig.\ref{fig-rot}).

In conclusion, we have found that the $\winf$ symmetry
allows for analytic description of
a non-trivial phenomena for quantum incompressible fluids in the
range of large edge excitations, characterized by
finite $R\to\infty$ limit.

Let us add some comments:
\begin{itemize}
\item
  The ansatz excitation (\ref{n-def}) is actually
  Girvin-MacDonald-Platzman original one \cite{gmp}, but our
  geometry is different, being the half-plane; thus, the edge roton in
  Fig.\ref{fig-rot} is unrelated to the bulk magneto-roton (for
  $\nu<1$).
\item
  The edge reconstruction phenomenon is qualitatively expected for
  any two-body interaction underlying the incompressible fluid state.
  For quantitative universal features, one should consider the density shape,
  to be discussed later.
\item
  Fig.\ref{fig-rot} manifestly shows non-linear behaviors occurring
  beyond the conformal invariant regime. As already said, the
  validity of our results on this domain follows
  the $\winf$ algebra (\ref{f-alg}) and the ground state conditions
  (\ref{hws-f}),
  that also hold for $k\sim R\to\infty$. Thus, it is a consistent
  extension, supported by the results of Section 2.2.
\item
  The expression (\ref{n-spec}) for $\eps({\bf n,p})$ can be integrated in
  terms of error functions ${\rm erf}(x)$
  without gaining in simplicity. This result was
  also obtained by using Laplace modes and the algebra (\ref{rho-alg}); once
  rewritten in terms of momenta, as in (\ref{f-def}), Eq. (\ref{n-spec}) is
  reproduced in the $R\to\infty$ limit (\ref{large-exc}).
 \end{itemize}


 \subsubsection{Spectrum for fractional filling}

 The derivation of the spectrum (\ref{n-def}) extends verbatim to $\nu=1/3$:
 the needed inputs, i.e. the bosonic form of the Hamiltonian (\ref{h-bose}),
 the algebra (\ref{f-alg}) and ground-state conditions (\ref{hws-f}) are
 unchanged, only the unit of length is redefined. From the analysis of Section
 3.4, we know that the correct scaling is given by $r^2\to r^2/\sqrt{m}$ for
 filling $\nu=1/m$. This fact can be accounted for by replacing
 $\ell^2\to\ell^2\sqrt{m}$ in all quantities requiring a length scale.  The
 expression of the spectrum (\ref{n-spec}),
 $\eps({\bf n,p})\sim {\bf n}^3 \ell^2$, acquires an overall factor $\sqrt{m}$
 and the same factor multiplies all arguments of exponentials and cosinuses.
 Alternatively, one can leave the formula (\ref{n-spec}) unchanged but
 remembers to rescale $\ell^2$. 

In table \ref{tab-min} we report our results for the
edge roton minimum ${\bf k}_o$ and  edge reconstruction distance 
$x_o$ (the latter is taken from density shape in Section 4.5),
and compare them with the values found in the literature \cite{edge-reco}
\cite{jain-rot}. 
Universality for these quantities is not to be expected, as said before.

\subsubsection{Bulk momentum dependence}

\begin{figure}
\begin{center}
\includegraphics[width=0.9\textwidth]{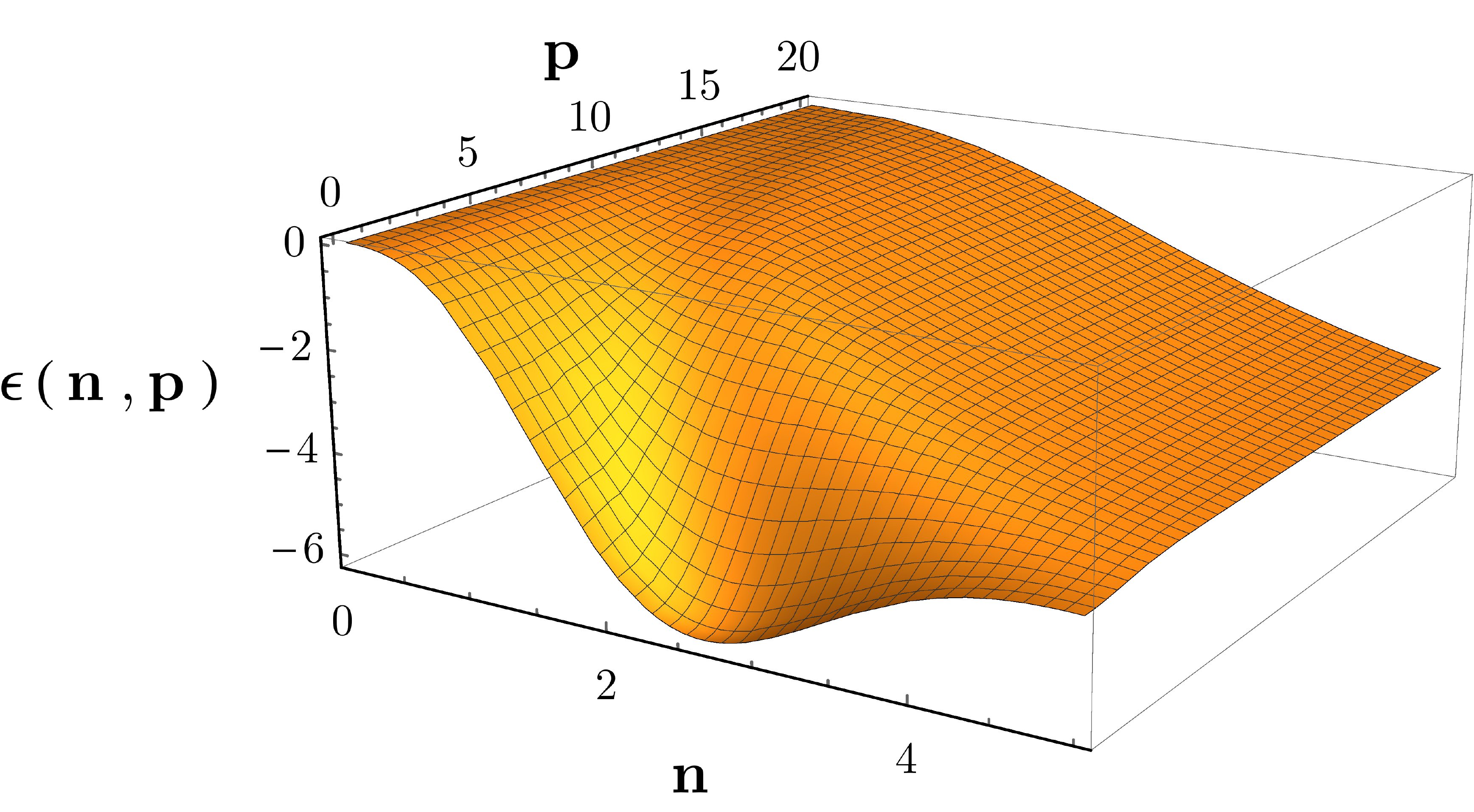}
\caption{Energy spectrum $\eps ({\bf n,p})$ of large neutral excitations 
  plotted as a function of ${\bf n}$ and ${\bf p }$ ($\nu=1$).}
\label{fig-pdep}
\end{center}
\end{figure}

The energy spectrum found in the previous Section also depends on the
momentum ${\bf p}$ orthogonal to the boundary, specifying the Fourier
component for the radial modulation of excitations. It is
found that $\eps({\bf n,p})$ is even in ${\bf p}$, as expected; it is
monotonically growing, eventually removing the minimum for
${\bf p}\gtrsim 10$ and asymptotically reaching a ${\bf p}$-independent
curve for large values (see Fig,\ref{fig-pdep}).
Actually, there is very little dependence on ${\bf p}$ outside the region
of the minimum.
The same qualitative behavior is found in the fractional case, 
only involving a rescaling of momenta.

The disappearance of the minimum can be interpreted as evidence that
the radial modulation spoils the edge reconstruction effect, i.e.
reduces the attractive exchange interaction.

The overall weak ${\bf p}$-dependence of the spectrum is a
consequence of the compact support of the excitation created by
$\r_{-n}({\bf p})|\W\rangle$. Actually, this is localized
on a region $|x|=O(1)$ and its Fourier transform is also localized
around $|{\bf p}|=O(1)$. In conclusion, the ansatz excitation
(\ref{n-ex}) analyzed in this Section
is not really appropriate for creating a bulk density fluctuation.

\begin{table}
\centering
\be 
\begin{array}{|l|l|l|l|l|}  
  \hline
 & {\bf k}_o (\nu=1) & x_o (\nu=1) &{\bf k}_o (\nu=1/3) & x_o (\nu=1/3)
\\
\hline
\winf {\ \rm symmetry}& 2.6 & 0.65 & 2.0& 0.85 \\
{\rm edge\ roton}  &  &  & 1.5 &   \\
{\rm edge\ reconstruction}  &  &\sim 2 & & \\
\hline
\end{array}
\nonumber
\ee
\caption{Minimum ${\bf k}_o$ of energy spectrum and positions $x_o$ obtained
  in our analysis versus results for edge reconstruction \cite{edge-reco} and
  edge roton \cite{jain-rot} (in units $\ell=1$).}
\label{tab-min}
\end{table}

\subsection{Spectrum of large charged excitations: bulk
  fluctuations}

In the following, we find the energy spectrum for the excitation
obtained by adding a big charge at the boundary, $Q={\bf t} R/m$,
${\bf t}=O(1)$.
This is the `large' version of the edge state described in 
Section 3.3, and is defined by:
\be
|\{{\bf n, p; t} \}\rangle=\r_{-n}({\mathbf p})|Q\rangle,
\qquad\quad Q=\frac{{\bf t} R}{m},\qquad \nu=\frac{1}{m} .
\label{q-def}\ee

We consider the expectation value, 
\be
  \eps({\bf n,p;t})=
  \frac{\langle\{{\bf n, p;t}\}|H |\{{\bf n, p;t}\}\rangle}
  {\langle\{{\bf n, p;t}\} |\{{\bf n, p;t}\}\rangle} ,
\label{q-def2}\ee
recalling that the Hamiltonian obeys $H|Q\rangle=0$ and is suitable
for computing excitation energies.
The derivation of the spectrum is obtained by using the $\winf$ algebra
and the highest-weight state conditions, as in previous Sections.
There are two additional terms w.r.t.  the earlier calculation, owing to
$\r_0({\bf p})|Q\rangle\neq0$, that are proportional to the expectation values
of the density and its square.

These expectation values are given by the
Casimir generating function (\ref{rho-q}) and (\ref{rho-frac}):
after replacing Laplace with Fourier variables according to (\ref{f-def2}),
one obtains the form,
\be
\frac{\langle Q|\r_0({\bf p})|Q\rangle}
{\langle Q|Q\rangle}
= \frac{1}{m}
\frac{e^{\frac{i{\bf pt}}{2}} -1} {i\frac{\bf p}{2R} }, \qquad\qquad
Q=\frac{{\bf t} R}{m}, \qquad \nu=\frac{1}{m}
\label{q-exp} \ee
bearing on the discussion in Section 3.4.
Note that this result can be simply obtained by Fourier transforming
the (classical) droplet of hight $1/m$ and sharp boundary.
As a matter of fact, this approximation is sufficient for the large
excitation regime.
 
The energy spectrum (\ref{q-def2}) is found to be:
\ba
&&\eps({\bf n,p;t}) =
\frac{ \sqrt{m}\; e^{-\frac{{\bf n}^2\sqrt{m}}{4} }} {\bf n}
\bigg\{
\nl
&&- 2
\int_0^{\bf n} d {\bf k}\; e^{-\frac{{\bf k}^2\sqrt{m}}{4}}
\left({\bf n-k}\right)
\left[{\bf k}^2 e^{\frac{{\bf n}^2\sqrt{m}}{4}}
  +\left({\bf n}^2-{\bf k}^2\right)
  \cos\left(\frac{{\bf p k}\sqrt{m}}{2} \right)   \right]
\nl
&&-  
\int_0^{\bf n} d  {\bf k}\int_0^{\bf n} d  {\bf k'}
e^{-\frac{({\bf k - k'})^2\sqrt{m}}{4}}
\left({\bf n}^2-({\bf k - k'})^2\right)
\cos\left(\frac{{\bf p (k-k')}\sqrt{m}}{2} \right)
\nl
&&+ 2 \frac{\sqrt{m}-1}{m}
\int_0^{\bf n} d  {\bf k}\int_0^{\frac{{\bf t }}{\sqrt{m}}} d  {\bf k'}
\left[
  e^{-\frac{({\bf k - k'})^2\sqrt{m}}{4}}
  \left( {\bf n}^2 - ({\bf k - k'})^2\right)
  \cos\left(\frac{{\bf p (k-k')}\sqrt{m}}{2} \right)
\right.
  \nl
&& \qquad\qquad\qquad\qquad\qquad\qquad
\quad -\ \left({\bf k'} \to -{\bf k'} \right)  \bigg] \bigg\} .
\label{q-spec}
\ea

\begin{figure}
\hbox{\hspace{-1cm}
\subfigure[\label{fig-qa}]{\includegraphics[scale=0.19]{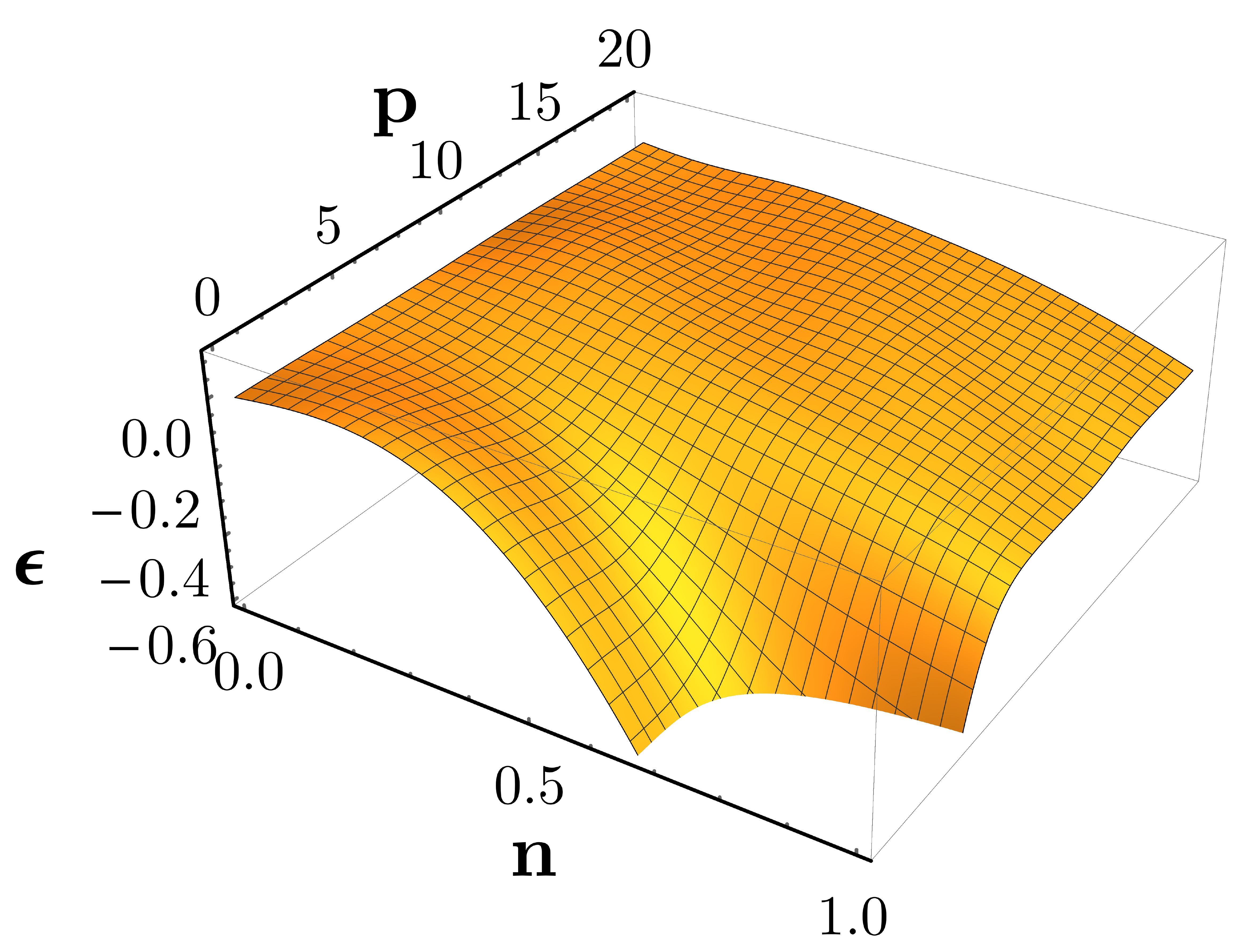}} 
\subfigure[\label{fig-qb}]{\includegraphics[scale=0.19]{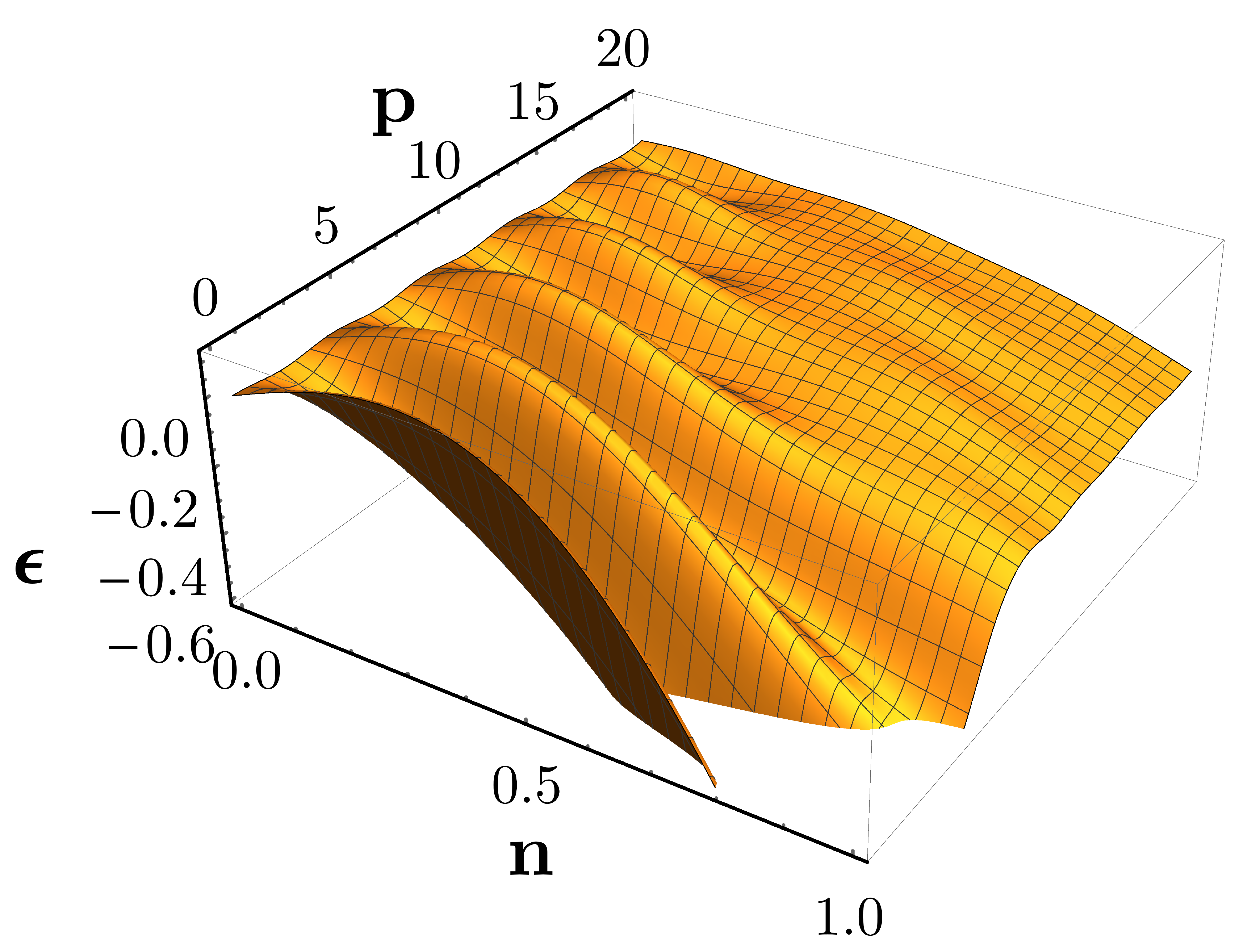}}}
\hbox{\hspace{-1cm}
\subfigure[\label{fig-qc}]{\includegraphics[scale=0.19]{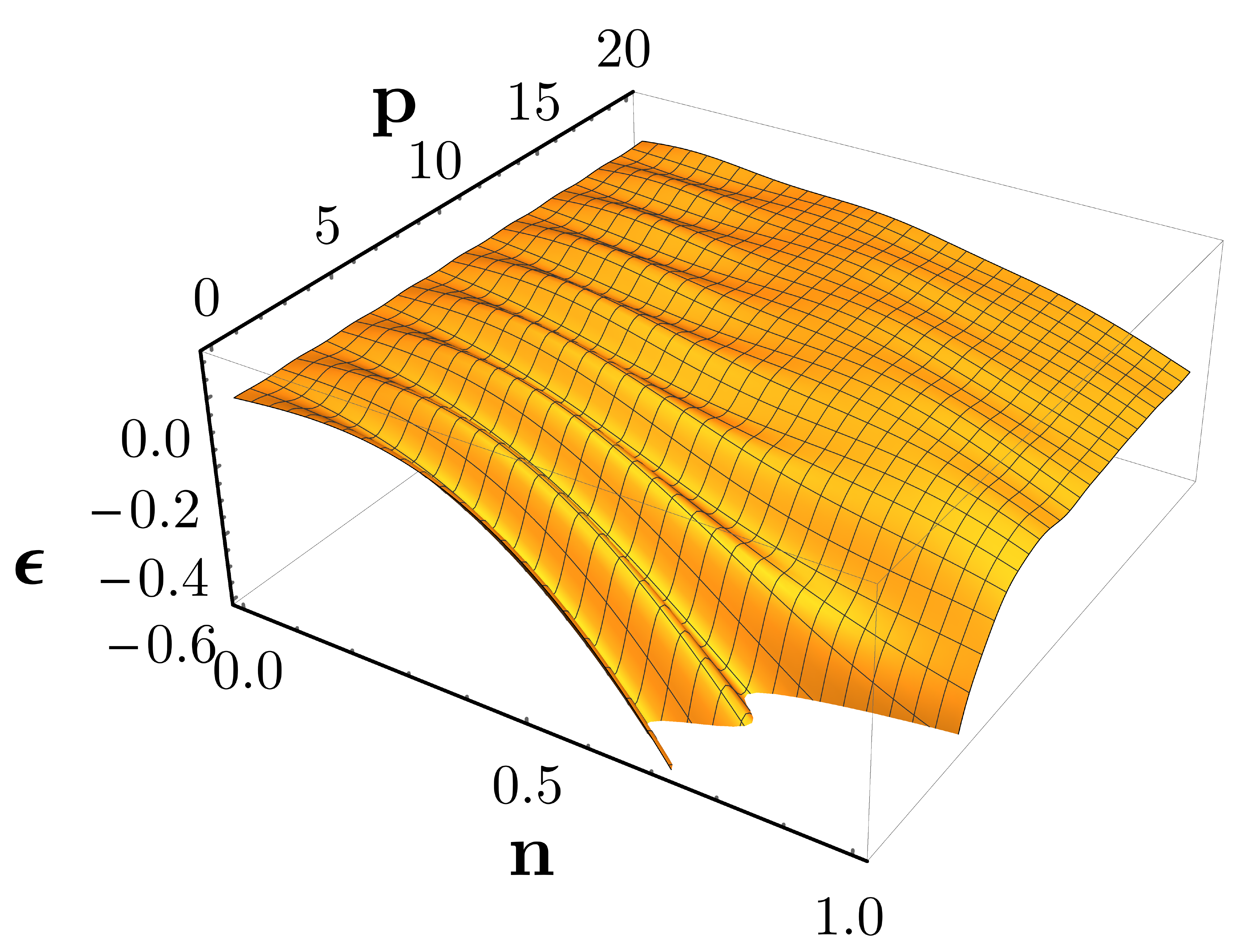}} 
\subfigure[\label{fig-qd}]{\includegraphics[scale=0.19]{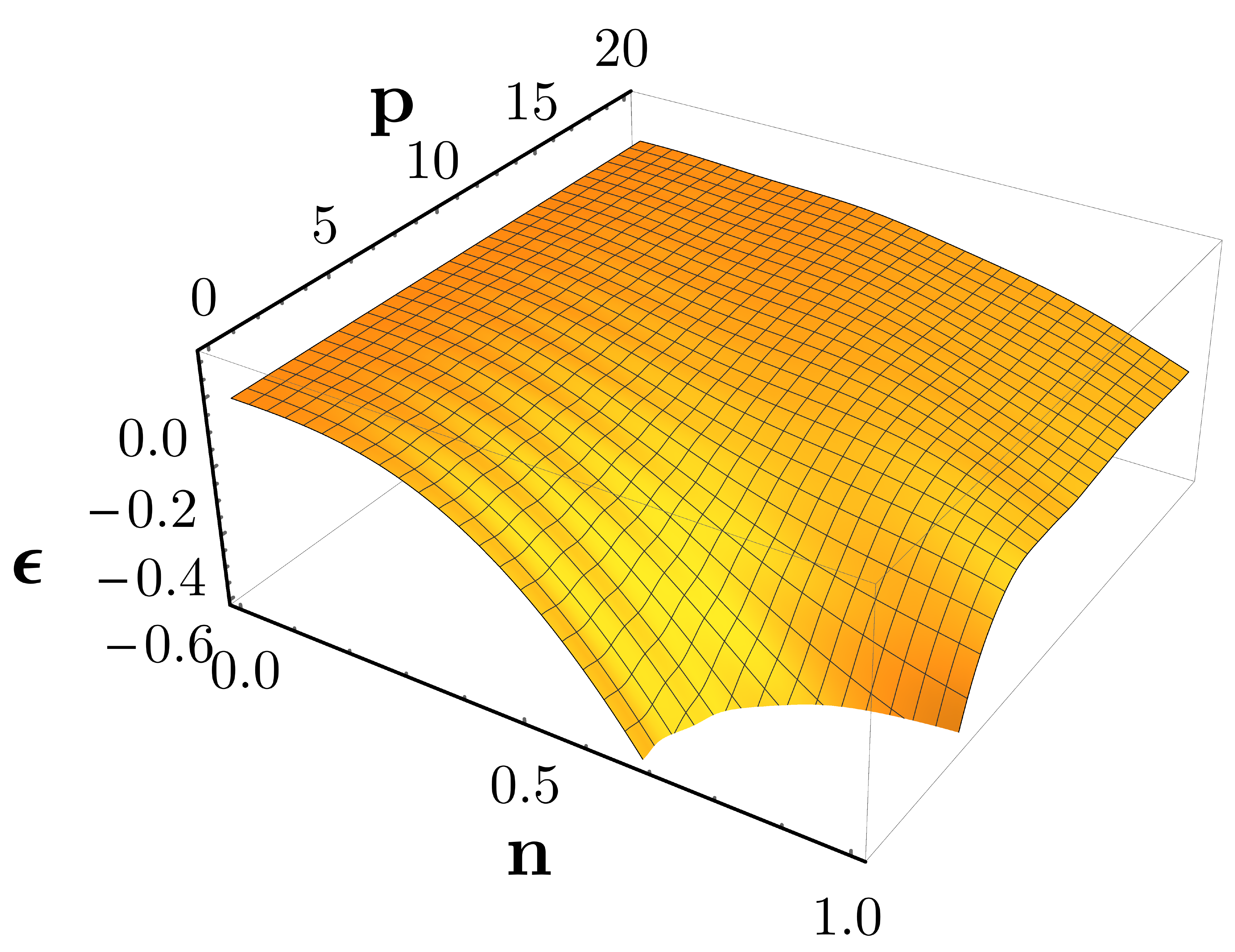}}}
\caption{Energy spectrum $\eps ({\bf n,p;t})$ for large charged
  excitations plotted as a function of ${\bf n}$ and ${\bf p }$ for:
  (a) ${\bf t }=1$; (b) ${\bf t }=3$ (c) ${\bf t }=5$; (d) ${\bf t }=7$
  ($\nu=1/3$).}
\label{fig-qdep}
\end{figure}

The first two terms in this expression are the same as in the neutral
$\nu=1$ case (\ref{n-spec}), suitably rescaled by
$\ell^2\to\ell^2\sqrt{m}$, as explained before. The third term is the
new charge-dependent part: it is of positive sign and non-vanishing
for fractional fillings. Actually, adding a charge to the
completely filled level does not change its shape, and only
corresponds to a redefinition of value of $R$. Thus, it is not
surprising that in this case the spectrum of excitations above
neutral and charged ground states are equal.

In the fractional case, the ${\bf p}$ dependence of the spectrum is
non-trivial: one sees dumped oscillations of period
$T_{\bf p}\sim 15/{\bf t}$ for values of ${\bf n}$ not too big
and charges ranging from ${\bf t}=1$ to ${\bf t}=7$
(see Fig. \ref{fig-qdep}).
This result shows that a finite deformation of the
density at the boundary due to charge accumulation determines a
response of the quantum incompressible fluid to bulk compressions.
The properties of these excitations will be further discussed
after computing the density profile in the next Section.

\subsection{The density profile for large excitations}

The shape of the fluctuations (\ref{q-def})
is given by the expectation value,
\be
\langle \r_0({\bf s})\rangle=
\frac{\langle \{{\bf n,p;t} \} |\r_0({\bf s}) |\{{\bf n,p;t} \}\rangle }
{\langle \{{\bf n,p;t} \} |\{{\bf n,p;t} \}\rangle},
\qquad\quad Q=\frac{{\bf t}R}{m},
\label{q-rho}\ee
where ${\bf s}$ is the momentum orthogonal to the boundary.
This expression can be again evaluated by using the $\winf$ algebra
in the Fourier basis (\ref{f-alg}). The definition of $\r_0({\bf s})$
needs the bulk prefactor (\ref{rho-lim1}), that should be regularized in the
map between Laplace and Fourier modes as follows:
\be
(1+\l)^{-R^2}\r_0(\l)\quad \to\quad |1+\l|^{-R^2}\r_0({\bf p})\sim
e^{-\frac{{\bf p}^2}{8}}\r_0({\bf p}),\qquad\ \
\l\sim -\frac{i{\bf p}}{2R} .
\label{rho-fact}\ee

The expression of $\langle \r_0({\bf s})\rangle$
is Fourier transformed back to coordinate space in terms
the variable $x=r-R$ near the edge, leading to
the result (see Appendix):
\ba
\!\!\!\!\!\!\!
\langle \d\r(x)\rangle&=&\frac{1}{\pi} \langle \r_0(x)\rangle
\nl
&=&
\frac{1}{m\pi}\bigg\{
\frac{1}{2}\left(
{\rm erf}\left(\frac{{\bf t}-2x}{\sqrt{2} m^{\frac{1}{4}}}\right)+
{\rm erf}\left(\frac{\sqrt{2} x}{ m^{\frac{1}{4}} }\right) \right)
\nl
&+&\frac{1}{\sqrt{2\pi }{\bf n}R }
\int_0^{\bf n} d {\bf k}\left[\left(
 e^{-\frac{2}{\sqrt{m}}\left(x-\frac{t-{\bf k}\sqrt{m}}{2} \right)^2}
   -e^{-\frac{2}{\sqrt{m}}\left(x-\frac{t+{\bf k}\sqrt{m}}{2}\right)^2}
   \right)
- ({\bf t}=0)\right]
\nl
&+&\frac{1}{\sqrt{2\pi}{\bf n}R }
\int_0^{\bf n} d {\bf k}\left(
 e^{-\frac{2}{\sqrt{m}}\left(x-\frac{{\bf k}\sqrt{m}}{2}\right)^2}
 -e^{-\frac{2}{\sqrt{m}}\left(x-\frac{({\bf k-n})\sqrt{m}}{2}\right)^2} \right)
\bigg\}.
\label{rho-bob}\ea
We note that the first and second terms in this expression
depend on the charge and
vanish for ${\bf t}=0$. The first part reduces to the edge expression
(\ref{rho-qfr}) for small fluctuations; the third part similarly matches
the particle-hole excitation (\ref{rho-net}) (see Section 3.4).
Note that all terms can be rewritten in terms of error functions,
but the integral forms can be more compact. 

\begin{figure}[h]
\begin{center}
{\includegraphics[scale=0.5]{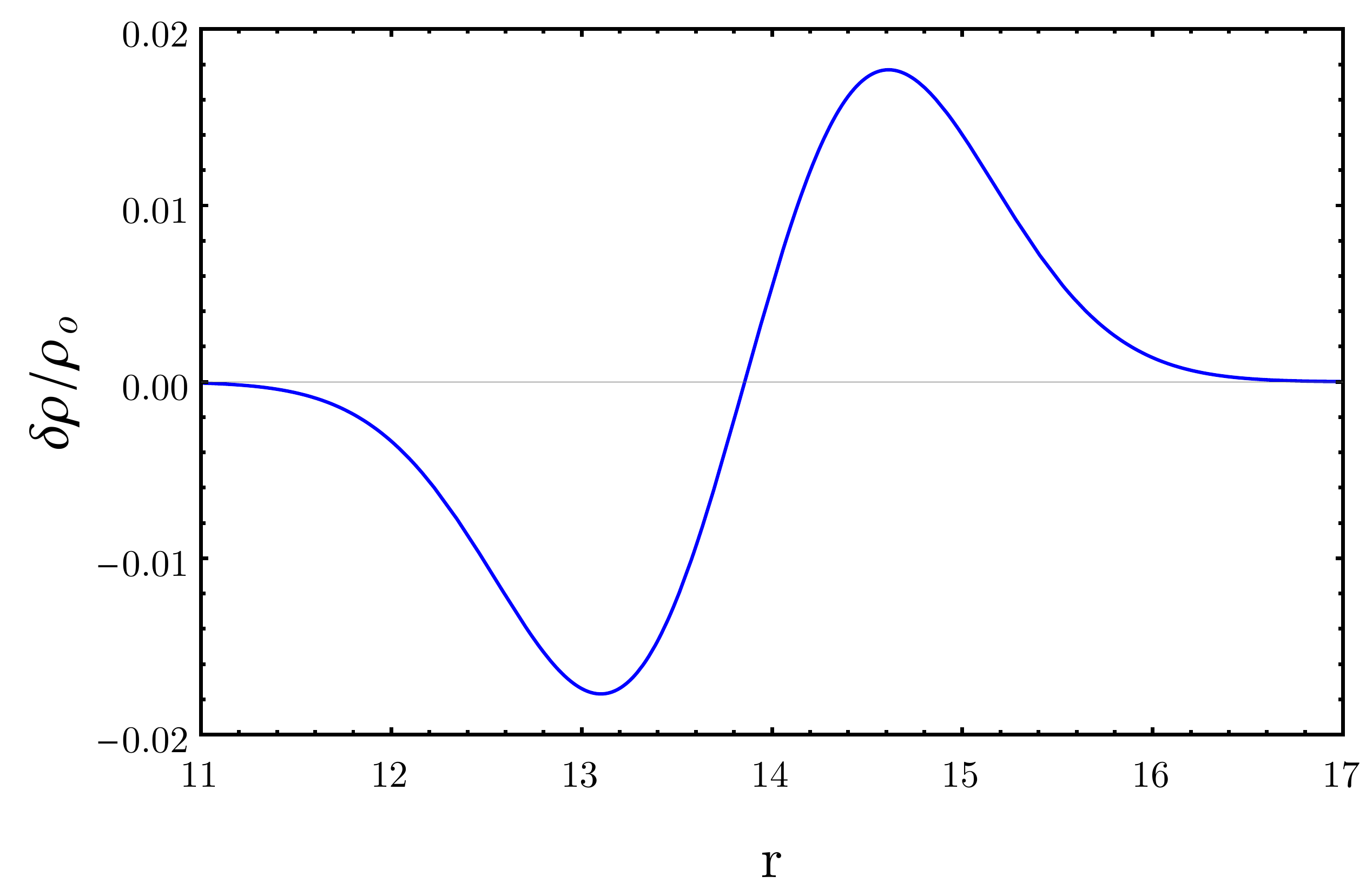}} 
\caption{Density profile of ${\bf n}=1$ large particle-hole excitation for
  $\nu=1/3$. The overall factor $1/R$ is fixed to the value of $N=64$ electrons
  for comparison with the small excitation in Fig. \ref{fig-frac}.}
\label{fig-er}
\end{center}
\end{figure}

\begin{figure}[h]
\begin{center}
{\includegraphics[scale=0.35]{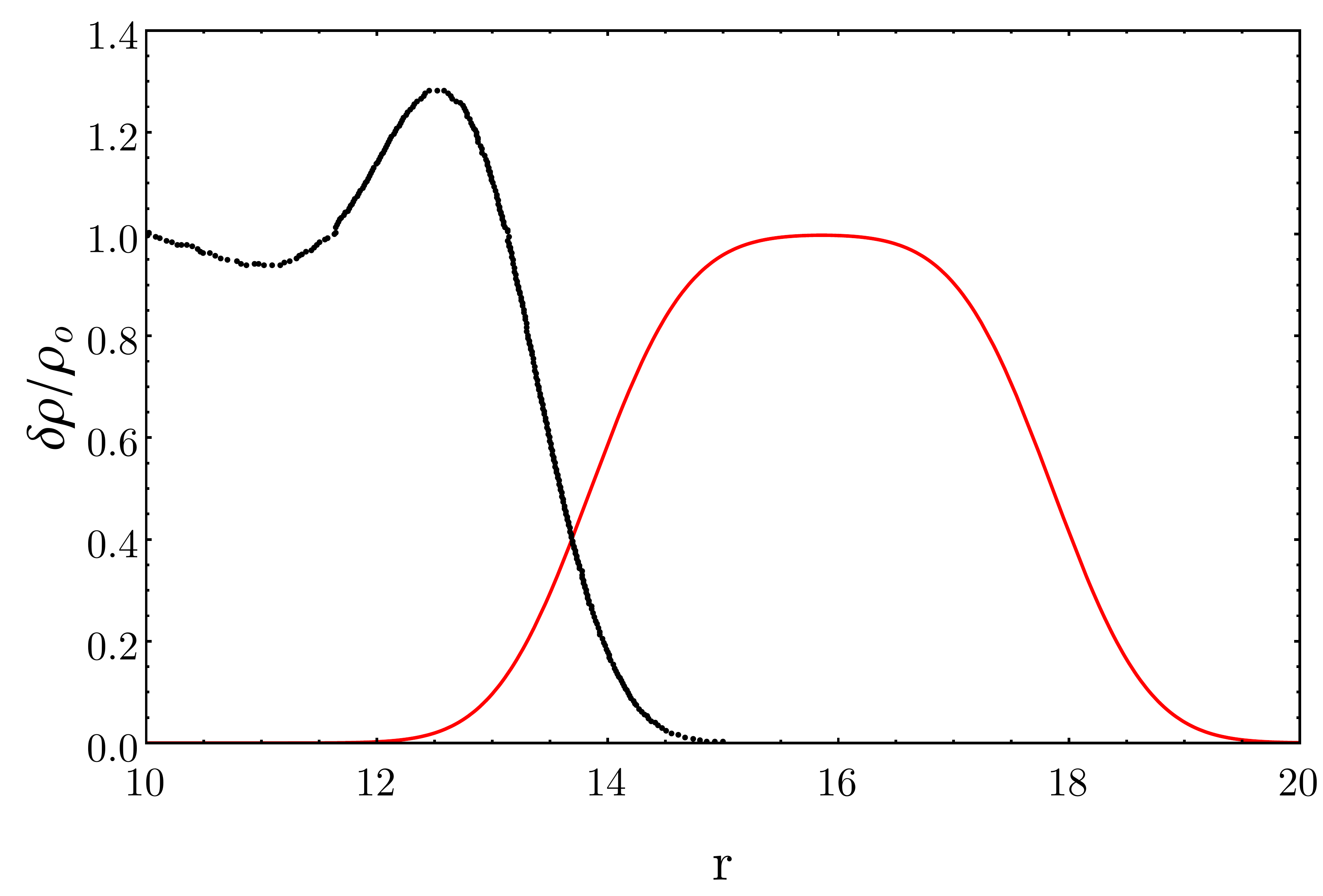}} 
\caption{Density profile of the ground state (black dots)
  \cite{laugh-num} and of the large charged excitation with
  ${\bf t}=8$ (red line) for $\nu=1/3$.  Both profiles are independent
  of the number $N$ of electrons, up to a shift on the $r$ axis
($N=64$ here).}
\label{fig-bob}
\end{center}
\end{figure}

Let us first discuss the case ${\bf t}=0$, given by the third term in
(\ref{rho-bob}).  Its profile looks like a regularized delta prime and
is of size $O(1/R)$ (see Fig.\ref{fig-er}).  Actually, this
fluctuation is too small to describe the edge reconstruction,
corresponding to a macroscopic displacement of matter, $\D \r=O(1)$:
this phenomenon is presumably given by the superposition of
$O(R)=O(\sqrt{N})$ such particle-hole excitations, corresponding to
the state $\prod_i^R\r_{-n_i}({\bf s})|\W\rangle$. The computation of
this energy is cumbersome, but at the semiclassical level it is
approximately linear in its components, leading to an accumulation of
the $\{n_i \}$ at the minimum of the spectrum (\ref{n-spec})
${\bf n}_o\sim 2.6/m^{1/4}$.
The density profile (\ref{rho-bob}) has a maximum at $x_o\sim{\bf n}/4$ for
large ${\bf n}$, but already accurate for
${\bf n}\sim {\bf n}_o$.
This determines the  splitting distance $x_o$ reported in Table \ref{tab-min}. 

In the case ${\bf t}\neq 0$, the first term in (\ref{rho-bob}) is
$\D\r=O(1)$, while the two other pieces are $O(1/R)$ and can be
neglected for $R\to\infty$.  The shape of density deformation is a
positive lump extending over several magnetic lengths. As shown in
Fig. \ref{fig-bob}, the large charge accumulated at the edge is
superposed to the ground state density terminating at $r\sim 15$
(for $N=64$ electrons).  Let
us incidentally note that this excitation cannot model the edge
reconstruction effect because it is attached to the ground state
density, not detached.

The large charged 
excitation has sufficient size and support to create a real bulk
fluctuation: as a consequence, it energy spectrum shows a non-trivial
dependence w.r.t. the bulk momentum ${\bf p}$, as seen in Fig.
\ref{fig-qdep}.
At present, we do not have enough elements for associating the dumped
fluctuations observed in the spectrum to specific bulk phenomena. The
large charged fluctuation is still localized near the edge, and thus it is
rather different from a low-energy density wave.
In our analysis, we considered ansatz excitations that are extension of
edge states: these are not the best suited for exploring
bulk physics.  We conclude that future investigations will require the
formulation of better ansatz states within the analytic setup
of this paper.

\section{Conclusions}

In this paper, we have shown that the $\winf$ symmetry of quantum
Hall incompressible fluids can be used to describe the radial shape of
edge excitations. We have identified two
regimes of small and large fluctuations. The first domain includes
the conformal invariant dynamics, whose energies
and amplitudes vanish in the thermodynamic limit $R\to\infty$.
While analytic and physical aspects were well understood in this case,
we have provided the additional information of density profiles.

The second domain of large fluctuations was uncharted. We have seen
that these excitations have a finite limit for $R\to\infty$ and
partially extend into the bulk. We have considered some ansatz states
and computed analytically their shape and energy spectrum for
short-range Haldane potential: a minimum at finite momentum has been
found that corresponds to the edge reconstruction phenomenon (and edge
roton). Energy oscillations with respect to the bulk momentum have
been observed in presence of a large charge accumulated at the
boundary. Several results presented in this work can be checked by
available numerical methods, as e.g. those of Ref.\cite{abanov}.

The $\winf$ symmetry approach has been formulated for $\nu=1$ and then
extended to fractional fillings $\nu=1/m$ by bosonization.  As a
consequence, results in the integer and fractional cases are very
similar, basically involving the (non-intuitive) magnetic-length
scaling $\ell^2\to\ell^2\sqrt{m}$.  This close correspondence is due
to the composite fermion map that is built-in in our approach, as
described in Section 2. 

A consequence of this map is that the unperturbed Laughlin ground
state cannot be compressed as much as the filled level: in technical
terms, this property is expressed by the highest-weight conditions
(\ref{hws-f}). Actually, we have seen that density
fluctuations originates by decompressions in the region $\D r=O(1)$ at
the boundary (see Section 4.4), and by adding boundary charge
(Section 4.5). Our approach might not account for the compressibility
of Laughlin plasma, as described e.g. by hydrodynamics \cite{hydro}.
However, it is believed that the composite-fermion theory captures
the low-energy  physics of Laughlin states; thus, our approach could
be sufficient in this respect.

In introducing and motivating our work, we asked some questions about
universality of bulk physics and the magneto-roton spectrum.  We actually
found a general framework for analytic studies of density fluctuations,
providing some universal features. However, we 
could not identify the excitation corresponding to the
magneto-roton. One open problem is that of finding appropriate ansatzes
for density waves, since the states considered in Section 4 have
finite support.  We hope that further
work will provide better approximate eigenstates in this
range: for instance, one could study the spectrum according to the
two-dimensional spin of excitations, since the leading low-energy
branch is believed to have spin two \cite{bimetric}
\cite{son-mr}.

Another line of development is the study of $\winf$ symmetry for
integer filling $\nu=n$ corresponding to several filled Landau levels.
Such extension is possible because the symmetry act independently
within each level \cite{cm}.  In this formulation, one can describe the Jain
states by bosonization, implementing the composite-fermion
correspondence, and get insight into the multiple branches of bulk
excitations known to appear in these cases \cite{mag-rot}.

\bigskip

{\bf Acknowledgments}

A.C. would like to thank C. A. Trugenberger and G. R. Zemba for sharing
their insight on the $\winf$ symmetry. We would like to thank  P. Wiegmann for
useful scientific exchanges. We also acknowledge the hospitality
by the G. Galilei Institute for Theoretical
Physics, Arcetri. This work is supported in part by the Italian
Ministery of Education, University and Research under the grant PRIN
2017 ``Low-dimensional quantum systems: theory, experiments and
simulations.''

\appendix


\section{Derivation of some formulas}

In this Appendix, we give some details of the calculations done in the
paper.

\subsection{Small particle-hole excitation}

The inverse Laplace transform of the expectation value
\eqref{rho-ph2} is:
\be
\label{Alap-inv-ph}
\langle \d \r_0(r^2)\rangle_{\{k\}}=- \int_{-i\infty}^{i\infty}
\frac{d\l}{2\pi i}\,e^{\l r^2}\frac{\left(1-\left(1+\l\right)^k\right)^2}
  {k\l\left(1+\l\right)^{R^2+k+1}}.
\ee
The integral is done by closing the path on the semiplane
Re$\left[\l\right]<0$ and calculating the residue at $\l=-1$
(recall that $R^2=N-1$ for $\nu=1$). This can be written as,
\be \langle \d
\r_0(r^2)\rangle_{\{k\}}=e^{-r^2}\frac{1}{\left(R^2+k\right)!}
\left.\left(\frac{d}{dt}\right)^{R^2+k}\left(e^{t
      r^2}\frac{\left(1-t^k\right)^2}{k\left(1-t\right)}\right)\right|_{t=0},
\ee
where we defined $t=\l+1$.  The following result is obtained:
\be
\label{Ap-h}
\langle \d\r_0(r^2)\rangle_{\{k\}}=
\frac{e^{-r^2}}{k}\frac{r^{2R^2}}{\G\left(R^2+1\right)}
\left\lbrace\left[\sum_{n=1}^{k}-\sum_{n=-k+1}^{0}\right]r^{2n}
  \frac{\G\left(R^2+1\right)}{\G\left(n+R^2+1\right)}\right\rbrace.
\ee
In the edge limit (\ref{edge-lim2}) $r=R+x$ and $R\to\infty$, the Stirling
approximations,
\ba
e^{-r^2}\frac{\left(r\right)^{2R^2}}{\G\left(R^2+1\right)}
&\sim& \frac{e^{-2x^2}}{\sqrt{2\pi}R}+O\left(\frac{1}{R^2}\right)\,,
\nl
\frac{\G\left(R^2+a\right)}{\G\left(R^2\right)}&\sim &R^{2a}
\left(1+\frac{a\left(a-1\right)}{2R^2}+O\left(\frac{1}{R^4}\right)\right),
\qquad a\sim O(1) ,
\label{Aapprox}
\ea
are used for obtaining the particle-hole edge excitation \eqref{rho-neu}.

\subsection{Small charged excitation }

The inverse Laplace transform of the expectation value
\eqref{rho-q} is:
\be
\label{Alap-inv-ch}
\langle \d \r_0(r^2)\rangle_{Q}=\int_{-i\infty}^{i\infty}
\frac{d\l}{2\pi i}\,e^{\l r^2}
  \frac{1-\left(1+\l\right)^{-Q}}{\l\left(1+\l\right)^{R^2+1}}.
\ee
We close again the path in the semiplane Re$[\l]<0$ and reduce to
the evaluation of the residue at $\l=-1$.  We get:
\be
\label{Arho-exp}
\langle \d \r_0(r^2)\rangle_{Q}=e^{-r^2}\frac{r^{2R^2}}{\G\left(R^2+1\right)}
\sum_{n=1}^{Q}r^{2n}\frac{\G\left(R^2+1\right)}{\G\left(n+R^2+1\right)}.
\ee
Upon using the edge approximation \eqref{Aapprox}, the expression
\eqref{rho-exp} is found for small charged excitations.

\subsection{Excitation of Laughlin states}

We evaluate the inverse Laplace transform of the expectation value
\eqref{rho-frac} for charged excitations of Laughlin state with $\n=1/m$.
The integrand in the following expression,
\be
\label{Alap-inv-chm}
\langle \d \r_0(r^2)\rangle_{Q}=\frac{e^{-\frac{r^2}{\sqrt{m}}}}{m}
\int_{1-i\infty}^{1+i\infty} \frac{dt}{2\pi
  i}\,e^{\frac{tr^2}{\sqrt{m}}}
\left(\frac{1-t^{-\frac{n}{\sqrt{m}}}}{\left(t-1\right)
    t^{\frac{R^2}{\sqrt{m}}}}\right),
\ee
has a branch cut starting
from the pole $t=1+\sqrt{m}\l=0$ along the negative real axes,
arg$(t)\in\left[-\pi,\pi\right]$. We consider the half `key-hole path'
of integration going around the singularity and the branch-cut.
The integral is rewritten:
\be
\label{Acut}
\langle \d \r_0(r^2)\rangle_{Q}
=\frac{e^{-\frac{r^2}{\sqrt{m}}}}{m}\left[
  \frac{\sin\left(\pi\left(\frac{R^2+n}{\sqrt{m}}\right)\right)}{\pi}
  \int_0^\infty dt\,\frac{e^{\frac{-t
        r^2}{\sqrt{m}}}}{\left(1+t\right)t^{\frac{R^2+n}{\sqrt{m}}}}-
  \left(n=0\right)\right].
\ee
The integral can be expressed in terms of the incomplete gamma function
$\G(a,z)$ and then analytically continued from negative to positive
$a$ values, also using the Euler's reflection formula for $\G(a)$.
The result is the expression (\ref{rho-gam}) given in the text.

The following integral form the incomplete gamma function,
\be
\G(a,z)=\int_{z}^{\infty}e^{-t}t^{a-1},
\label{Ainc-g}
\ee
is useful for computing the edge limit \eqref{edge-lim2}.
We consider its derivative, expand it in $x=r-R$ for large $R$ and
then integrate it back. The result \eqref{rho-qfr} follows:
\be
\langle \d \r_0(r^2)\rangle_{Q}=\int_{-\infty}^{x}\de_{y}\langle
\d \r_0((R+y)^2)\rangle_{Q}=\frac{e^{-\frac{2x^2}{\sqrt{m}}
  }}{m^\frac{1}{4}\sqrt{2\pi} R}
\frac{n}{m}+O\left(\frac{1}{R^2}\right).
\ee

\subsection{Large excitations $k=O(R)$}

The Laplace transformed density \eqref{lap-alg} is now
approximated within the ranges of coordinates and momenta given in
\eqref{large-exc}. The Laplace modes for $k={\bf k}R$ become:
\be
\r_k(\l)= \sum_{j'=-R^2}^\infty \frac{1}{(1+\l)^{R^2+{\bf
      k}R/2+j'+1}} \frac{\G\left(R^2+{\bf k}R/2+j'+1
  \right)}{\left(\G(R^2+j'+1)\G(R^2+{\bf k}R+j'+1)\right)^{1/2}}
c^\dag_{j'} c_{j'+k}.
\label{Alap-alg}
\ee
Upon using the improved asymptotic of the gamma function (cf.
(\ref{Aapprox}),
\be
\frac{\G\left(R^2+{\bf k}R+a\right)}{\G\left(R^2\right)}
\sim R^{2{\bf k}R+2a}e^{{\bf k}^2/2}\left(1+O\left(\frac{1}{R}\right)\right),
\ee
one finds:
\be
\r_{{\bf k}R}(\l)= e^{-{\bf k}^2/8}\frac{1}{\left(1+\l\right)^{R^2+\mu+1} }
\sum_{j'=-R^2}^\infty (1+\l)^{\mu-j'-{\bf k}R
/2} c_{j'}^\dag c_{j'+{\bf k}R},
\qquad R\to\infty .
\ee
Note the presence of the exponential prefactor
$\exp{\left(-k^2/8R^2\right)}$, with $k={\bf k}R$, as introduced in
\eqref{large-rho}.

\subsection{Two-body Hamiltonian}

Let us consider the matrix element \eqref{h-gauss} and expand the
field operators \eqref{field-op} in Fock space.
The expression to be evaluated is:
\be
H_t=-\sum_{k,r,s\in\mathbb{Z}} M_t\left(k,r,s\right)c^\dag_s c_{s-k}
c^\dag_r c_{r+k},
\ee
\be
M_t\left(k,r,s\right)=-2\int_0^\infty dr_1\int_{0}^\infty dr_2 r_1r_2\,
    I_k(2tr_1r_2)e^{-\left(1+t\right)\left(r_1^2+r_2^2\right)}
    \frac{r_1^{2s+k}r_2^{2r-k}}{\sqrt{\left(s+k\right)!s!\left(r-k\right)!r!}},
\label{Atwo-body}
\ee
where $I_k(z)$ is a Bessel function.
The angular momentum indices $\{r,s,k\}$ and the radial coordinates are
shifted and expanded in large excitation limit \eqref{large-exc}.
The suitable asymptotic of the Bessel function is the following:
\be
I_k\left(z\right)\sim \frac{e^{z-\frac{k^2}{2z}}}{\sqrt{2\pi z}}
    \left(1+O\left(\frac{1}{z}\right)+ O\left(\frac{k^4}{z^3}\right) \right),
    \qquad z\to\infty, \quad k=O(\sqrt{z}).
\ee
The $r_1, r_2$ integrals in (\ref{Atwo-body}) are computed by saddle
point method following the same steps as in Ref.\cite{ctz-dyna},
but the results are now valid in the range of large excitations.
We find the result:
\be
\label{Aht}
H_t=\frac{2}{Rt^2}e^{-\frac{k^2+\left(r-s+k\right)^2}{4R^2}}
\left(k^2-\left(r-s-k\right)^2\right)+O\left(\frac{1}{t^3}\right),
\ee
where we also antisymmetrized $M_t\left(r,s,k\right)$ respect to the
fermion exchange $\{r\to s,\,s\to r,\,k\to -k\}$.
Finally, taking the $t\to\infty$ limit \eqref{del-h} for \eqref{Aht},
the result \eqref{h-hald} for Haldane potential is obtained.

\subsection{Energy spectrum of large excitations}

The expectation value of the Hamiltonian (\ref{q-def2}) involves the
expression:
\be
\langle Q| \r_{n}({\bf -p})\r_{-k}({\bf -q}) \r_{k}({\bf q}) 
  \r_{-n}({\bf p})|Q\rangle , \qquad\quad k>0,
\ee
that can be rewritten as,
\be
\langle Q|\left[ \r_{n}({\bf -p}),\r_{-k}({\bf -q})\right]
\left[ \r_{k}({\bf q}),   \r_{-n}({\bf p})\right]Q \rangle.
\ee
The evaluation of the commutators with the help of the algebra (\ref{f-alg})
gives three terms:
\ba 
&&A_{n,-k}(-{\bf p},-{\bf q})A_{k,-n}({\bf p},{\bf q})
\langle Q| \r_{n-k}({\bf -p-q})\r_{k-n}({\bf p+q})|Q\rangle 
\nl
&&+\d_{n,k}A_{n,-n}(-{\bf p},-{\bf q})B_{n}({\bf p+q})
\langle Q|\r_{0}({\bf p+q})|Q\rangle+h.c.
\nl
&&+\d_{n,k}B_{n}({\bf-p-q})B_{n}({\bf p+q}) \langle Q|Q\rangle.
\label{Acomm}
\ea
For vanishing charge $Q=0$, i.e. for neutral excitations analyzed
in Section 4.4, the term in the second line of (\ref{Acomm}) vanishes,
while the first part can be replaced by the commutator
 $\left[\r_{n-k}({\bf -p-q}),\r_{k-n}({\bf p+q})\right]$, for $k< n$.
 In the charged case, the non-vanishing second term is
 proportional to $\langle\r_{0}\rangle$, while  the expression in the first
 line acquires the $k=n$ contribution proportional to $\langle|\r_0|^2\rangle$.
 Thus, there are two terms for $Q=0$ and four for $Q\neq 0$.

 All contributions are summed over $k>0$ in the Hamiltonian
 and contain other 
 summations in the $B_n$ terms. Discrete summations
 can be approximated by integrals for large excitations
$n={\bf n}R$:
\be
\frac{1}{R}\sum_{k=1}^{{\bf n}R}=\int_{0}^{{\bf n}}d{\bf k}, \qquad
k= {\bf k} R.
\ee
Note that in this limit the large contributions from the density expectation
value (\ref{q-exp}) $\langle \r_0\rangle=O(R)$ can never be neglected.
With these ingredients, the energies of 
neutral and charged large excitations, respectively \eqref{n-spec} and 
\eqref{q-spec}, are obtained.

The corresponding expressions in the fractional case $\nu=1/m$ are obtained
by scaling $\ell^2\to\ell^2\sqrt{m}$ in all dimensionful quantities,
keeping in mind that the dimensionless $\langle \r_0\rangle$ acquires
an extra factor $1/\sqrt{m}$ from (\ref{rho-match}).

\subsection{Large density profiles}

The expectation value \eqref{q-rho} can be written as:
\be
\langle
\r_0({\bf s})\rangle=\frac{\langle Q|\left[ \r_{n}({\bf
      -p}),\r_{0}({\bf s}) \r_{-n}({\bf p})\right]|Q\rangle}
{\langle Q|\left[\r_{n}({\bf -p}),\r_{-n}({\bf p})\right]|Q\rangle},
\ee
being $n>0$. The use of the algebra \eqref{f-alg} leads to three terms:
\ba
\!\!\! \! \! \langle
\r_0({\bf s})\rangle &=&
\frac{\langle Q|\r_{0}({\bf s})|Q\rangle}{\langle Q|Q\rangle}
\nl
&+&
\frac{A_{n,0}({\bf -p},{\bf s})A_{n,-n}({\bf -p+s , p})}{B_n(0)}
\frac{\langle Q|\r_{0}({\bf s})|Q\rangle}{\langle Q|Q\rangle}
\nl
&+& \frac{A_{n,0}({\bf -p},{\bf s})B_{n}({\bf s})}{B_n(0)}.
\ea
For large charge $Q=tR/m$ and angular momentum $n={\bf n}R$, the first
term is of order $O(1)$ while the two others are $O(1/R)$.
The inverse Fourier transform of $\langle \r_0({\bf s})\rangle$
can be formulated directly w.r.t. the boundary variable $x=r-R$,
after including the bulk prefactor (\ref{rho-fact}),
as follows:
\be
\langle \r_0(x)\rangle =\frac{1}{4\pi R}\int dx\;
e^{-i{\bf s}x -{\bf s}^2/8} \langle \r_0 ({\bf s})\rangle .
\ee
The result \eqref{rho-bob} is obtained.


\end{document}